\documentstyle[psfig,12pt]{article}

\begin{document}


\title{\Large
Row-switched states in two-dimensional \\
underdamped Josephson junction arrays}

\author{
$\begin{array}{cc}
\mbox{\large Mauricio Barahona} &
\mbox{\large Shinya Watanabe$^*$} \\
\mbox{\it Ginzton Laboratory} &
\mbox{\it Center for Chaos \& Turbulence Studies} \\
\mbox{\it Stanford University} &
\mbox{\it Niels Bohr Institute} \\
\mbox{\it Stanford, CA 94305} &
\mbox{\it Blegdamsvej 17, Copenhagen} \\
\mbox{\it mauricio@loki.stanford.edu} &
\mbox{\it DK-2100, Denmark}
\end{array}$
}

\date{\normalsize submitted, October, 1997}

\maketitle

\begin{abstract}

When magnetic flux moves across layered or 
granular superconductor structures,
the passage of vortices can take place 
along channels which develop finite voltage,
while the rest of the material remains in the zero-voltage state.
We present analytical studies of an example of 
such mixed dynamics: the row-switched (RS) states
in underdamped two-dimensional Josephson arrays,
driven by a uniform DC current under external magnetic field
but neglecting self-fields.
The governing equations are cast into 
a compact differential-algebraic system
which describes the dynamics of an assembly of Josephson oscillators
coupled through the mesh current.
We carry out a formal perturbation expansion,
and obtain the DC and AC spatial distributions
of the junction phases and induced circulating currents.
We also estimate the interval of the driving current
in which a given RS state is stable.
All these analytical predictions compare well with our numerics.
We then combine these results to deduce the parameter region
(in the damping coefficient versus magnetic field plane)
where RS states can exist.\\

\end{abstract}


\section{Introduction}
\label{sec:intro}

\begin{figure}[tbp]
\centerline{\psfig{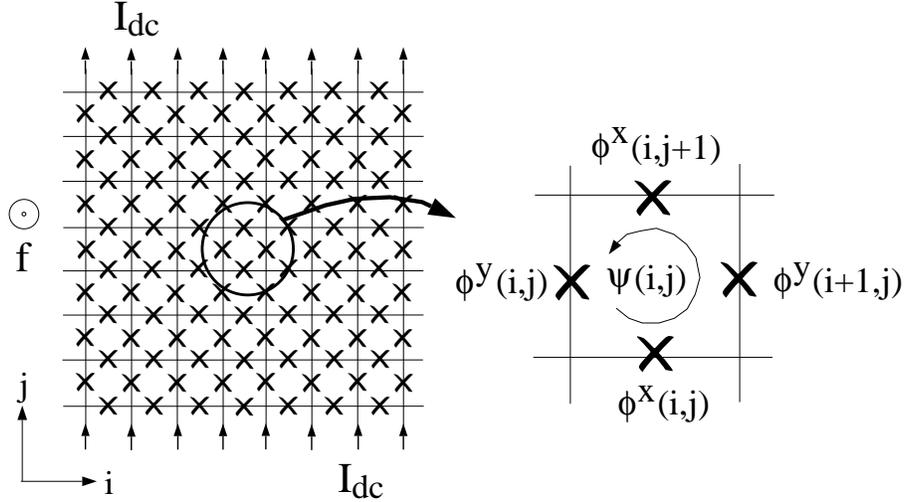}}
\vspace{5mm}
\caption{2D Josephson junction array consisting of $N^x=7$ rows
and $N^y=7$ columns of square cells.
The cell at $(i,j)$ is shown enlarged.
Each junction is described by a gauge-invariant phase difference: 
$\phi^x$ for the junctions on the horizontal edges, 
and $\phi^y$ for the vertical junctions. 
A uniform DC bias current 
$I_{\mbox{\scriptsize dc}}$ is injected 
into every node on the bottom edge
and extracted from the top. The left and right sides are open boundaries.
The mesh current $\psi$ denotes the deviation of the current distribution 
from a uniform current flow in the vertical direction.
A uniform magnetic field $f$, 
in units of the flux quantum $\Phi_0$,
is applied transversally to the plane of the array.
}
\label{fig:array}
\end{figure}

Two-dimensional (2D) arrays of Josephson junctions serve as
``controlled laboratories'' to investigate fundamental questions 
such as phase transitions~\cite{herrephasetran}, 
vortex propagation and 
interaction~\cite{geigenlobb,vortexparticle,dominguez,josevortex}, 
phase- and frequency-locking of coupled
oscillators~\cite{wiesen1,landsberg,filawiesen,oppenlaender},
and spatio-temporal pattern formation
and chaos~\cite{marino,spatiotempchaos},
among others~\cite{trieste95}.
A standard circuit geometry is a rectangular array driven 
by a DC current uniformly injected from the 
bottom and extracted from the top in the presence of an applied field
(Fig.~\ref{fig:array}).
Their technological promise as 
high-frequency oscillators~\cite{benz,shorted2D,duwel} depends critically on
achieving tunable, highly nonlinear, coherent oscillations
of the collection of junctions.
However, to the chagrin of engineers
(and to the curiosity of dynamicists)
such coherent oscillations are not easy 
to obtain \cite{wiesen1,duwel,1Dseries}.
Instead, the arrays frequently break up into incoherent substructures,
and deliver output voltages with small AC amplitudes.

\begin{figure}[tbp]
\centerline{\psfig{file=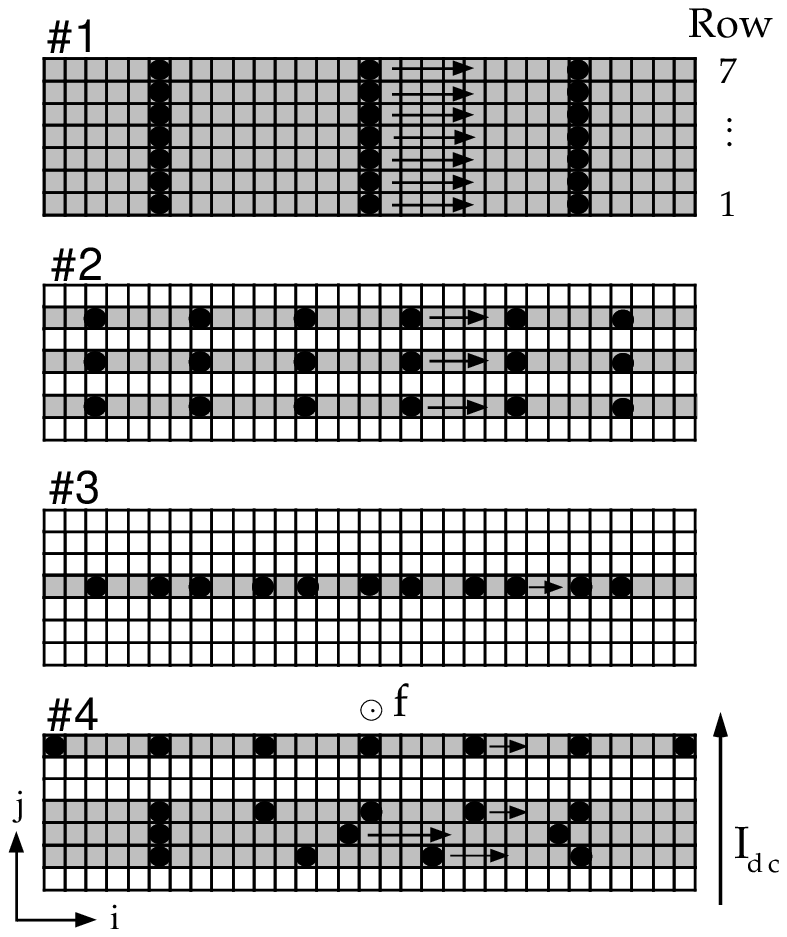,height=5in}}
\vspace{5mm}
\caption{
Four snapshots of RS states
in arrays of $N^x=31$ columns by $N^y=7$ rows.
Two types of rows are observed:
quiescent (Q) rows (in white)
across which there are zero DC voltage drops,
and switched (S) rows (shaded areas)
across which there are finite DC voltage drops.
Black dots denote topological vortices,
defined in Sec.~\protect{\ref{sec:formulation}}.
They are (roughly) equally spaced in the $S$ rows of the
symmetric Patterns 1--3,
but the spacing can change from row to row
in asymmetric patterns such as Pattern 4.
Correspondingly, their propagation speed (represented by the
length of the arrows)
may change from an $S$ row to another within a pattern.
In Patterns 1 and 2,
the vortices move in phase,
even when the $S$ rows are separated by $Q$ rows.
These patterns are numerically generated 
using $\Gamma =0.2$, $f=0.1$ and 
$I_{\mbox{\scriptsize dc}}=0.6$ for Patterns 1,2 and 4, and
$I_{\mbox{\scriptsize dc}}=0.5$ in the case of Pattern 3.
Thus, Patterns 1,2 and 4 correspond to {\it coexisting}
dynamical attractors of the system.
}
\label{fig:patterns}
\end{figure}

A striking example of such dynamical states with spatial structure is 
provided by the {\em row-switched} (RS) states
found in underdamped 2D arrays of square cells~\cite{experimrs}.
As the bias current $I_{\mbox{\scriptsize dc}}$ is ramped up, 
the DC current--voltage ($I$--$V$) 
characteristic of the array displays a succession of discontinuous
jumps between ohmic branches of increasing resistance until, 
eventually, the normal resistive branch is reached.
It was also noticed that the branches are equally spaced in voltage.
These observations suggested a row-switching scenario,
in which each of the jumps corresponds to 
all the junctions in an individual row suddenly switching
from the superconducting to the normal state,
thus increasing the voltage across the array by a fixed amount.
In the RS states, the array then consists of superconducting 
and normal rows, coexisting to form striped patterns as in the
four examples shown in Fig.~\ref{fig:patterns}.
In other words, the magnetic flux moves across the array along
certain rows (channels) where a finite voltage develops, while
the rest of the system remains in the zero-voltage state.
This row-switching picture was later explicitly confirmed by measuring 
voltages across individual rows \cite{ETthesis,ETatASC}, 
and by direct imaging of the array~\cite{lachenmann}.
Moreover, attempts were made \cite{ETthesis} 
to determine the sequence of row-switching as the bias current was varied,
and showed that the order and critical currents
at which rows become normal are irregular and history-dependent.

The row-switching phenomenon is robust and still appears when the
underlying lattice structure is changed.
In arrays of triangular cells it has been observed
both  experimentally~\cite{herreballistic} 
and in simulations~\cite{stroud}. 
The explanation of row-switching could also be relevant to other systems.
For instance, similar channeling of magnetic flux
has been seen in continuous superconductors~\cite{channeling}.
In addition, 
the hundreds of resistive steps which appear
in the $I$--$V$ characteristics of high-Tc superconductors~\cite{highTcSC}
have been taken as an indication 
of the layered weak-link structure in those materials.

The experiments on 2D arrays of square cells have revealed
that RS states appear only when the junctions are sufficiently 
underdamped~\cite{experimrs,ETatASC,lachenmann,recentlachenmann}.
Otherwise, the $I$--$V$ characteristics 
present an extended region of flux-flow 
leading to the ohmic branch of the entire array.
In addition, RS states are only observed for 
sufficiently small applied magnetic fields.
If the field is too large, 
there are no individual RS steps; rather, 
one giant step emerges~\cite{ETatASC,ETthesis} in the $I$--$V$.
The origin of this giant step has been attributed 
to the interaction of self-fields 
with a coherent array oscillation 
in the form of a dynamical checkerboard pattern~\cite{ETstep}.
To understand such transitions between
coherent and localized states,
it is important to study the RS states
and determine their current and phase distributions
as well as the parameter regime for their appearance.

Much of the previous theoretical work on 2D arrays has consisted of numerical 
simulations~\cite{sawada,chung,majhofer,losalamos,dominguez,stroud,geigenlobb,joelstatic,joelrs,ETthesis,recentlachenmann}
which reproduce the measurements reasonably well.
Several of them have discussed RS states
briefly~\cite{majhofer,geigenlobb},
while three others have treated in depth
the row-switching phenomenon~\cite{stroud,joelrs,ETthesis}.
Their main conclusion is the characterization of the two distinct types 
of rows found in the experiments:
(1) {\em switched} rows (``S'' rows), 
across which there is a finite DC voltage,
and (2) {\em quiescent} rows (``Q'' rows), 
across which there is no DC voltage drop.
The simulations show that
the junctions $\phi^y$ in the vertical branches of the $S$ rows are 
in the normal resistive state (rapidly rotating)
whereas those in the $Q$ rows are nearly superconducting (stationary).
Nevertheless,
as we show, the junctions in the $Q$ rows are still
oscillating, which causes finite AC voltage drops
and associated losses.
This is why we hesitate to call the $Q$ rows ``superconducting''.

The numerical investigations have also studied
the sequence of row-switching as the bias current is varied.
Even in the absence of temperature and disorder,
the observed patterns and the order of their appearance
are found \cite{stroud,ETthesis} to depend on factors such 
as the initial condition,
how currents are varied, the magnetic field (both externally applied
and self-induced), among others.
This is a clear indication that multiple attractors coexist 
for identical parameter values.
(Indeed, Patterns 1, 2, and 4 in Fig.~\ref{fig:patterns}
are found using the same set of parameters 
while Pattern 3 is obtained
for a slightly smaller bias current.)
When inhomogeneity is included,
it becomes even harder to predict which row will switch next,
except to conjecture that it will occur 
at the ``weakest part'' of the array~\cite{stroud}.
Phillips {\em et al.}~\cite{joelrs} have studied the vortex patterns
in detail when inductances are included. When self-fields are small, 
the $S$ rows appear to be globally phase-locked
even if they are far apart, separated by $Q$ rows in between.
This means that topological vortices in the $S$ rows
appear to propagate together,
just as seen in Pattern 2 in Fig.~\ref{fig:patterns}.
However, for generic asymmetric patterns, such as Pattern 4,
vortices do not move together. 
Stronger self-fields are also found \cite{joelrs}
to break this phase coherence.

Compared to the numerous experimental and numerical studies,
analytical results are much scarcer for 2D arrays.
As far as we are aware,
previous authors have focused on the simplest solution,
namely, when the whole array is on the normal branch of the $I$--$V$
curve (Pattern 1 in Fig.~\ref{fig:patterns}).
This can be interpreted as the special RS state
when all the rows have become normal; that is, the 
``completely row-switched'' solution.
These studies have concentrated on explaining
the global phase-locking mechanism
needed for oscillator applications.
The complete RS state is found to be only neutrally stable
under zero magnetic field~\cite{wiesen1,petraglia}
(which implies that rows are decoupled),
whereas a non-zero field induces inter-row locking.
This inter- and intra-row coupling mechanisms have been studied
by several methods:
isolating two cells in the array~\cite{larsen,landsberg},
perturbation methods~\cite{landsberg}, 
and harmonic balance~\cite{marino,filawiesen}.
However, those results are not directly applicable to generic RS states
since the complete RS state has no $Q$ rows
and it extends to any large bias current values for any damping.
Instead, the (generic) RS states
exhibit non-trivial structures and 
exist only in a certain parameter regime.

In this paper, we study analytically the RS states
and test our predictions
against numerical integrations of the system.
First, we cast the governing equations and the boundary conditions 
in the  mesh formalism to ease the analytical procedure
(Section~\ref{sec:formulation}).
In this notation, the system can be viewed as an array of coupled oscillators 
in which the junction phases $\phi$ (the pendulum-like oscillators)
are coupled via the mesh currents $\psi$ (the
current distribution in the array).
The coupling arises from the flux quantization condition.
We neglect self-field effects in the equations,
thus reducing the parameters of the system to three:
the bias current $I_{\mbox{\scriptsize dc}}$,
the junction damping coefficient $\Gamma$,
and the magnetic field $f$.
In this way, many properties of the RS states can be explained
without undue complications.
We also discuss the notion of vorticity in these discrete arrays.

We use primarily four examples (depicted in Fig.~\ref{fig:patterns})
in order to illustrate and test our results.
It is convenient to label each RS pattern by the {\em set} S
of its switched rows.
Therefore, Patterns 1 to 4 are labeled as:
$S=\{1,2,3,4,5,6,7\}$, $S=\{2,4,6\}$,
$S=\{4\}$, and $S=\{2,3,4,7\}$, respectively.
We also define an {\em S region} to be a set of {\em contiguous} $S$ rows.
For example, Pattern 4 in Fig.~\ref{fig:patterns} has
two $S$ regions, one with three rows 2--4 and 
another with a single row 7.
Similarly, a {\em Q region} is a set of contiguous $Q$ rows.

A formal perturbation expansion in the high-frequency limit~\cite{likharev}
is used to analyze the governing equations (Section~\ref{sec:analysis}).
We assume that the RS states are time-periodic solutions
in which some junctions whirl
(i.e.,  the $\phi^y$'s in the $S$ rows are running oscillators),
and all the other junctions librate 
(i.e., the $\phi^y$'s in the $Q$ rows and all $\phi^x$ are nearly stationary).
Although the expansion is made systematic so that
higher-order corrections could be obtained,
we show that most of the features
of the RS states can be accounted for by the leading order.
(The only unresolved main feature is the phase-locking
between $S$ rows.)
To the zeroth order, 
we obtain two systems of algebraic equations: one for 
the DC, and another for the AC components of the phases and currents.
The DC system is nonlinear (thus difficult to solve); however, we obtain
bulk approximations which work well far from the edges.
On the other hand, the AC components obey the linear discrete Poisson
equation with forcing from the DC solution; 
therefore, they can be readily obtained
{\it once} the DC solution is known.

The bulk approximation determines analytically 
the DC and AC distributions of
currents and phases for any given RS pattern. 
The first important result is that the DC current flows 
uniformly in the $S$ rows, but
circulating currents are induced in the $Q$ regions.
These strongly affect the spatial wave numbers of the $S$ rows
(also calculated analytically), thus explaining why the spacing and 
speed of propagation 
of the fluxoids in the $S$ rows varies from pattern to pattern, 
and even from row to row within a pattern (Fig.~\ref{fig:patterns}).
In Section~\ref{sec:numerics} we test these findings numerically
with good agreement.

Another main conclusion from the leading order analysis is that 
the presence of $S$ regions breaks the array into a collection 
of $Q$ regions that are {\em decoupled} from each other,
{\em as far as the DC equations are concerned}.
The $Q$ regions are, however, still 
weakly coupled through the AC component.
Thus, for example, the existence of the switched row 4 in Pattern 3 
produces two $31 \times 3$ quasi-disjoint $Q$ regions
which only interact weakly.
This picture proves useful because it reduces the problem of 
approximating the {\em dynamical} RS states of the array 
to obtaining the {\em static} states of its (smaller) 
constitutive $Q$ regions.

Indeed, this physical picture has further implications 
for the {\it stability} of the RS patterns (Section~\ref{sec:existence}).
As explained above,
each RS state is only observed in an interval of 
the bias current, which depends on the magnetic field and damping.
We show that the upper current limit of this interval 
is well predicted by the {\it depinning} current of the largest $Q$ region.
This means that the RS state ceases to exist when the flux
penetrates any of the Q regions which, in absence of irregularities, is
usually the largest one in the array.
For example, Pattern 3 cannot hold beyond the current value
at which a static state of a $31 \times 3$ array depins.
Along the same lines, we also argue that 
the largest upper current of any RS state will correspond to 
RS patterns whose largest $Q$ region is a single row,
such as Pattern 2.
However, the depinning currents are independent
of the junction damping, whereas the RS states are found
only in underdamped arrays. This indicates that the
lower current limit also plays an important physical role. 
A crude approximation for this lower limit is 
the retrapping current of a single junction which
does depend on the damping.
Combining these criteria we are then able to calculate the region of 
the parameter plane of the magnetic field $f$ versus the
damping parameter $\Gamma$ where RS states are {\it not} possible.  
Throughout Section~\ref{sec:existence} we present additional numerical
evidence which support these criteria.

\newpage
\section{Formulation}
\label{sec:formulation}

There are two equivalent ways of formulating the governing equations 
of the system: the node and mesh formulations. 
The node formulation is easier for simple geometries but
it becomes cumbersome and impractical 
for two-dimensional arrays when inductances are included. 
Thus, we follow the previous
literature~\cite{sawada,majhofer,dominguez,joelstatic,joelrs,ETthesis},
and derive a compact description of the arrays
in the mesh formulation.
In particular, we follow closely
Phillips {\em et al.}~\cite{joelstatic,joelrs}
and Tr\'{\i}as \cite{ETthesis},
with a few changes.
Although this formalism was originally developed to ease numerical simulations,
it is well suited for analytical work.

\subsection{Governing equations}

Our description of the array shown in Fig.~\ref{fig:array}
assumes several simplifications.
First, we neglect thermal fluctuations (i.e., zero temperature), and
we consider all junctions identical (i.e., no disorder). Second, 
we describe our basic circuit unit, a single Josephson junction,
by the resistively and capacitively shunted junction (RCSJ) model.
In this standard model, a junction driven by a current $I^b$ 
is represented by an equivalent circuit 
of three channels in parallel with a capacitance $C$, a resistance $R$, and 
a tunnel junction with the critical current $I_c$. 
As a result, its state variable $\phi$
(the gauge-invariant phase difference across the junction)
is governed by
\begin{equation}
    {\cal N}[\phi] \equiv \ddot{\phi}+ \Gamma \dot{\phi} + \sin \phi = I^b
\label{dimensionalRSJ}
\end{equation}
where the nonlinear operator ${\cal N}$ returns the total current
through the device. In (\ref{dimensionalRSJ}) the
current is normalized by $I_c$, whereas 
time is expressed in units of the inverse of the plasma frequency 
$\omega_p^{-1} = (\Phi_0 C/ 2 \pi I_c)^{1/2}$. In addition, 
$\Gamma= \beta_c^{-1/2} = (\Phi_0/2 \pi I_c R^2 C)^{1/2}$ is
the damping, with $\beta_c$ the McCumber parameter.
Also, $\Phi_0$ is the quantum of magnetic flux. 
The instantaneous voltage across the junction is
given by the Josephson voltage-phase relation:
\begin{equation}
   V(t) = \Gamma \dot{\phi} ,
\label{dimensionalvoltagephase}
\end{equation}
where the voltage is normalized by $I_c R$.
Thus, a single junction is analogous to 
a damped driven mechanical pendulum,
and its voltage corresponds to the rotation frequency
of the pendulum~\cite{likharev,steve,scbooks}.

When several junctions are interconnected to form a network,
like the one depicted in Figure~\ref{fig:array}, 
the current distribution must fulfill Kirchhoff's current law. 
This results in coupling among the junctions.
It is convenient to decompose each branch current 
into an external and a deviation current:
\begin{equation}
  I^b = I_{\mbox{\scriptsize ext}} + I_{\mbox{\scriptsize dev}} .
\label{currentdecomposition}
\end{equation}
The external current $I_{\mbox{\scriptsize ext}}$ is
{\em chosen} such that it
satisfies current conservation at all nodes,
including external sources and sinks.
In general, it can be spatially non-uniform or time-dependent.
However, in our case, the steady bias current
$I_{\mbox{\scriptsize dc}}$ is 
injected (extracted) at the nodes along the bottom (top) edges.
Therefore, our choice~\cite{choicecaveat} for $I_{\mbox{\scriptsize ext}}$
is the {\it stationary uniform vertical flow}, in which
$I_{\mbox{\scriptsize ext}}=I_{\mbox{\scriptsize dc}}$
on every vertical branch of the circuit (for all $t$), 
and $I_{\mbox{\scriptsize ext}}=0$ on every horizontal bond.

The deviation $I_{\mbox{\scriptsize dev}}$ from the external flow
must be divergence-free since current sources and sinks
have already been incorporated.
Therefore, there is a stream function (or mesh current) $\psi$
at each cell whose discrete curl determines $I_{\mbox{\scriptsize dev}}$
in the $x$ and $y$ directions:
\begin{eqnarray}
  I_{\mbox{\scriptsize dev}}^x(i,j) = \psi(i,j)-\psi(i,j-1) 
\label{curlofmeshcurrent1}\\
  I_{\mbox{\scriptsize dev}}^y(i,j) = -( \psi(i,j)-\psi(i-1,j) ).
\label{curlofmeshcurrent2}
\end{eqnarray}
(In the rest of this paper we will not write 
time-dependences explicitly when they are obvious, such as here.)

In order to ensure that these relations hold also at 
the edges of the array,
we define artificial {\em boundary cells} 
which have either the horizontal index $i=0$ or $N^x+1$, 
or the vertical index $j=0$ or $N^y+1$.
This yields the boundary conditions of the problem:
\begin{equation}
  \psi(i,j) = 0 \quad 
      \mbox{if $i=0,N^x+1$ \hspace{.1in} or \hspace{.1in} if $j=0,N^y+1$.}
\label{BC}
\end{equation}
This condition is equivalent to ``grounding'' the value of $\psi$ 
outside the array.

Combining~(\ref{dimensionalRSJ})--(\ref{curlofmeshcurrent2}), 
we obtain the first two sets of governing equations
\begin{eqnarray}
  {\cal N}[\phi^x(i,j)]  & = & \psi(i,j)-\psi(i,j-1)
\label{goveqn1} \\
  {\cal N}[\phi^y(i,j)] & = & I_{\mbox{\scriptsize dc}} 
  - (\psi(i,j)-\psi(i-1,j))
\label{goveqn2} 
\end{eqnarray}
where ${\cal N}$ was defined in (\ref{dimensionalRSJ}).

The other source of intrinsic coupling between the junctions 
is due to a macroscopic quantum constraint: 
the flux quantization condition around a cell.
Each corner of a cell is a superconducting island
described by a well-defined phase. 
Calculating the phase change around cell $(i,j)$ 
yields the third and final set of equations of the system
\begin{equation}
  ( \phi^y(i+1,j) - \phi^y(i,j) ) -
  ( \phi^x(i,j+1) - \phi^x(i,j) )
  + 2 \pi \frac{\Phi(i,j)}{\Phi_0} = 2\pi n(i,j)
\label{FQ}
\end{equation}
for $i=1,\dots,N^x$ and $j=1,\dots,N^y$,
where $\Phi(i,j)$ is the total magnetic field
penetrating the cell.
The winding numbers $n(i,j)$ are a set of integers 
that arise because the island phases are only
defined up to multiples of $2\pi$.
The initial condition sets $n(i,j)$ which remain constant
as long as the array is kept superconducting.
However, without loss of generality,
all $n(i,j)$ can be set to zero.
Suppose they are not zero;
then we can redefine the junction phases as
\begin{eqnarray}
  \phi^x(i,j) & \rightarrow & \phi^x(i,j) \nonumber \\ 
  \phi^y(i,j) - 2 \pi \sum_{k=1}^{i-1} n(k,j) & 
  \rightarrow & \phi^{y}(i,j)
\label{angleconversion}
\end{eqnarray}
such that (\ref{FQ}) is unchanged
except, now, {\it all} $n(i,j) \equiv 0$.
Crucially, both currents and voltages are invariant 
under this redefinition of the phases 
since adding integer multiples of $2\pi$ to $\phi^y$
changes neither $\sin \phi^y$ nor $\dot{\phi}^y$.
This means that the dynamics and measurements remain {\it identical}
for any combination of integers $n(i,j)$,
and we do not need to be concerned with their initial values.
Similarly, if the magnetic field were controllable independently on each cell,
adding an integer number of flux quanta $\Phi_0$ 
into any cell would not change
the measured $I$--$V$ characteristics,
at least within this model.
This is simply the array analogue of the 2-junction SQUID,
whose dependence on the penetrating field is also $\Phi_0$-periodic.
Because of this periodicity in the magnetic field, 
the topological vortex must be defined
differently in 2D arrays and in continuous superconductors,
as we will discuss at the end of this section.

The total magnetic field in~(\ref{FQ}) can be decomposed into two parts:
\begin{equation}
  \Phi(i,j) = \Phi_{\mbox{\scriptsize ext}} +
  \Phi_{\mbox{\scriptsize ind}}(i,j) .
\label{totalflux}
\end{equation}
The first term is produced by 
the external field applied perpendicularly to the plane of the array,
which we assume to be constant and uniform.
It is usually parametrized as a dimensionless frustration $f$
normalized to the flux quantum:
\begin{equation}
  f = \Phi_{\mbox{\scriptsize ext}}/ \Phi_0 , 
\label{frustration}
\end{equation}
such that, in terms of $f$, the period of the external field is unity.
The second term, the induced field,  
can be written generally as the sum of all the contributions
from the branch currents
\begin{equation}
  \Phi_{\mbox{\scriptsize ind}}(i,j) = 
  \sum_n \sum_k L^b_{n,k} I^b_k
\label{inducedfield}
\end{equation}
where $k$ runs through all the branches of the circuit
while $n$ corresponds to the four edges of cell $(i,j)$.
The branch inductances $L^b_{n,k}$ are purely geometric constants
determined from the circuit.~\cite{dominguez,joelstatic}

\subsection{Matrix-vector notation}

The above equations can be cast into
a compact matrix-vector notation~\cite{joelstatic}.
For a $N^x \times N^y$ array, all branch variables 
(e.g., currents $I^b$, voltages $V$, and phases $\phi$) 
can be written as vectors
of dimension equal to the number of branches,
i.e.\ $(N^x+1)N^y+N^x(N^y+1)$. 
Thus, for instance, the vector $\phi$ consists of 
all the phases $\phi^x$ and $\phi^y$.
On the other hand, variables defined at cells
(e.g., the mesh current $\psi$ and the
induced flux $\Phi_{\mbox{\scriptsize ind}}$) 
form vectors of dimension $N^x N^y$.
These two groups of vectors are connected via 
a branch-to-cell connectivity matrix~\cite{strang} 
$M$ which takes a directed sum as we loop around a cell:
\begin{eqnarray}
  M \phi(i,j) & = & 
  [\phi^y(i+1,j)-\phi^y(i,j)] \nonumber \\
  & - & [\phi^x(i,j+1)-\phi^x(i,j)].
\label{phaseloopsum}
\end{eqnarray}
More mathematically, this operator takes the discrete curl 
of $\phi$ around every cell $(i,j)$.
Conversely, the discrete curl of the cell variables is obtained by
applying the transpose cell-to-branch matrix $M^T$.

Using this notation, the total flux~(\ref{totalflux}) can be  written as
\begin{equation}
  \Phi = \Phi_0 f + M L^b I^b .
\label{vectortotalfield}
\end{equation}
where $L^b$ is the branch inductance matrix, and $f$ is a 
constant {\it vector}.

Moreover,
(\ref{curlofmeshcurrent1}),(\ref{curlofmeshcurrent2}) can now be written
simply as
\begin{equation}
  I_{\mbox{\scriptsize dev}} = M^T \psi,
\label{celldifference}
\end{equation}
and (\ref{goveqn1}),(\ref{goveqn2}) become
\begin{equation}
  {\cal N}[\phi] = I_{\mbox{\scriptsize ext}} + M^T \psi
\label{vectorRSJ}
\end{equation}
where ${\cal N}$ is operated component-wise and 
the vector $I_{\mbox{\scriptsize ext}}$ has components
that are zero on the horizontal edges and 
$I_{\mbox{\scriptsize dc}}$ on the vertical edges, as defined by our
choice of uniform vertical flow.

Finally, we can use~(\ref{phaseloopsum}) and (\ref{vectortotalfield}) to
recast the flux quantization condition~(\ref{FQ}) as
\begin{equation}
  M \phi + 2 \pi f +  
 \frac{1}{\lambda_\perp} (L^m \psi + M L^b I_{\mbox{\scriptsize ext}}) = 0
\label{vectorFQ}
\end{equation}
where components of $L^b$ are normalized to $\mu_0 p$,
$p$ is the lattice constant,
$\lambda_\perp = \Phi_0/2\pi I_c \mu_0 p$ is the dimensionless
penetration depth, 
the {\it mesh} inductance matrix is defined as
\begin{equation}
  L^m=M L^b M^T ,
\label{inductancerelation}
\end{equation}
and we have set $n(i,j) \equiv 0$.

To summarize,
the governing equations (\ref{vectorRSJ}) and (\ref{vectorFQ})
form a closed differential-algebraic system for $\phi$ and $\psi$,
with parameters $f$, $\Gamma$,
$I_{\mbox{\scriptsize dc}}$, $\lambda_\perp$,
and the coefficient matrix $L^b$.
This form of the system is compact and intuitive.
It can be seen as a coupled-oscillator system
in which the ``oscillators'' $\phi$ are driven 
by the ``coupling field'' $\psi$ in (\ref{vectorRSJ}).
In return, the oscillators collectively feed back onto the field 
in (\ref{vectorFQ}).
This picture suggests the following integration steps~\cite{joelrs,ETthesis}:
first, given $\phi$ at some time $t$, 
solve (\ref{vectorFQ}) for $L^m \psi$;
then, invert the matrix $L^m$,
together with the boundary conditions (\ref{BC}),
to determine the field $\psi$.
This gives us the ``drive'' on the right hand side of (\ref{vectorRSJ}),
which is used to update the oscillators $\phi$ in time.

We conclude the general formulation by pointing out that the equations 
(\ref{vectorRSJ}), (\ref{vectorFQ}) possess two simple 
symmetries~\cite{majhofer}.
If we find a solution $(\phi(i,j,t),\psi(i,j,t))$ 
for $f$ and $I_{\mbox{\scriptsize dc}}$,
then $(-\phi(i,j,t),-\psi(i,j,t))$ is a solution of the system
for $-f$ and $-I_{\mbox{\scriptsize dc}}$,
the other parameters being the same.
Similarly, $(-\phi(-i,-j,t),\psi(-i,-j,t))$ is also a solution 
for $f$ and $-I_{\mbox{\scriptsize dc}}$
(since $M$ is changed to $-M$
due to the reversal of the spatial coordinates).
Therefore, we only have to study the quadrant $f \geq 0$
and $I_{\mbox{\scriptsize dc}} \geq 0$.
Together with the unit periodicity in $f$, the 
frustration can be further restricted to $0 \leq f < 1/2$
without loss of generality.
In the rest of this article, 
expressions such as ``large $f$'' and ``small $f$''
are used within this interval.

\subsection{No-inductance approximation}

Computing the {\it full} equations~(\ref{vectorRSJ}),(\ref{vectorFQ})
quickly becomes a heavy task as the system size increases.
In previous numerical studies, these computational limitations have 
been circumvented either by using acceleration 
schemes~\cite{dominguez,joelstatic}
when the inductance effects are of interest {\it per se},
or by ``truncating'' the matrix $L^m$ (i.e. neglecting some of its 
components). Three truncations~\cite{dominguez,joelstatic,joelrs,ETthesis} 
are often used:
no-, self-, and nearest-neighbor-inductances.
Self-inductance neglects the inter-cell magnetic coupling 
by keeping only the diagonal 
components of $L^m$ (which then becomes trivially
invertible). 
Nearest-neighbor-inductance includes, in addition, 
magnetic coupling between neighboring cells.
An important remark is that
not only the mesh inductance $L^m$ 
but also the vector $M L^b I_{\mbox{\scriptsize ext}}$
must be provided in order to complete the system,
and the choice of $I_{\mbox{\scriptsize ext}}$
may affect the results
when $L^b$ is truncated~\cite{dominguez,ETthesis}. 
(In contrast, the choice of $I_{\mbox{\scriptsize ext}}$ is
unrestricted if the full inductance matrix is used.)
Truncating the system in a physically consistent manner is a subtle issue,
and, for simplicity, we shall assume no inductance in this article.

In contrast to what one might guess from its name,
the no-inductance approximation does {\em not} eliminate the 
inter-cell coupling.
Counterintuitively, it leads to an even longer-range coupling 
than the self- and nearest-neighbor- truncations.
The no-inductance approximation sets $L^b=0$, thus $L^m=0$.
The flux quantization condition~(\ref{vectorFQ}) is then just
\begin{equation}
  M \phi + 2 \pi f = 0 .
\label{goveqn3}
\end{equation}
The same equation can also be obtained in the limit
$\lambda_\perp=\infty$ for any $L^b$, which
allows the no-inductance limit to be approached
from the inductive system continuously. It is important to note that
the condition~(\ref{goveqn3}) is now a {\em constraint} on $\phi$,
which must be satisfied at {\it all} times. 
The discrepancies between $M \phi$ and $-2 \pi f$ cannot
be filled by {\it locally} adjusting the induced field,
which was possible when the inductive terms were present.
This leads to a global coupling of the junctions
over the whole domain.
To see the coupling mechanism provided by (\ref{goveqn3}),
operate the loop-sum $M$ on (\ref{vectorRSJ}).
From the left hand side we obtain
\[
  M {\cal N} (\phi) (i,j) = M \ddot{\phi} + \Gamma M \dot{\phi}
  + M[\sin \phi]
\]
but the first two terms vanish since (\ref{goveqn3}) 
must hold identically.
From the right hand side, we obtain
\[
  M (I_{\mbox{\scriptsize ext}} + M^T \psi)
  = M I_{\mbox{\scriptsize ext}} + M M^T \psi 
  = M I_{\mbox{\scriptsize ext}} - \Delta \psi
\]
where we have introduced the discrete Laplacian
\begin{eqnarray}
  \Delta \psi (i,j) & \equiv &
  ( \psi(i,j+1)+\psi(i,j-1) \nonumber \\
  & + & \psi(i+1,j)+\psi(i-1,j) )-4 \psi(i,j) .
\label{discreteLaplacian}
\end{eqnarray}
For the stationary uniform flow $I_{\mbox{\scriptsize ext}}$,
the term $M I_{\mbox{\scriptsize ext}}=0$.
Thus, we arrive at a discrete Poisson equation
\begin{equation}
  \Delta \psi = -M[\sin \phi]
\label{psiPoisson}
\end{equation}
in which the distribution of the mesh current is dependent
on all the junctions in the array.

Equations (\ref{vectorRSJ}) and (\ref{psiPoisson})
constitute the governing equations for the no-inductance case,
and can be integrated in the same manner as before.
Provided that the initial condition 
satisfies the constraint (\ref{goveqn3}) and 
its time derivative $M \dot{\phi}=0$, 
(\ref{goveqn3}) is satisfied for all $t$.
An immediate advantage of the no-inductance approximation is that the
sweep of the parameter space is greatly simplified since 
the number of parameters has been reduced to three: $f$, $\Gamma$, and
$I_{\mbox{\scriptsize dc}}$. 

\subsection{Vorticities in 2D arrays}
\label{sec:vorticities}

Before closing this section, we consider now the concept of
vorticity in these arrays. 
Analogously to incompressible planar fluid flows, 
we can define a vorticity 
by taking the curl (by applying $M$) 
of the ``velocity'' field
which, in our case, corresponds \cite{strang} to the branch current $I^b$.
This {\em current vorticity}
\begin{equation}
  \Omega = M I^b = M (I_{\mbox{\scriptsize ext}} + M^T \psi)
  = M I_{\mbox{\scriptsize ext}} - \Delta \psi
\label{defcurrentvorticity}
\end{equation}
measures how currents whirl, and can take any real values.
(For the uniform vertical external flow
the contribution $M I_{\mbox{\scriptsize ext}}$ vanishes.)

In contrast, the notion of a {\em topological vortex} (or charge) 
is commonly used in Josephson arrays 
in analogy to continous superconductors.
In Type-II superconductors, the vortices would correspond to the 
integer winding numbers $n(i,j)$ in the flux quantization 
condition~(\ref{FQ}).
But, as we showed above, the $n(i,j)$ are irrelevant in the arrays.
Therefore, an alternative, less physical definition
for the {\it topological vorticity} has 
been used:~\cite{geigenlobb,dominguez,joelrs,ETthesis} 
\begin{equation}
  \zeta = \frac{1}{2\pi} M (\widehat{\phi} - \phi) .
\label{deftopologicalvorticity}
\end{equation}
Here, $\widehat{\phi}$ denotes restriction of the components
of the phase vector $\phi$ within $[-\pi,\pi)$.
The value of $\zeta$ at each cell
takes only integer values  (typically 0 or $\pm 1$) and 
jumps discontinuously as the system evolves in time. 
In effect, this definition detects   
when one of the four junctions in a cell rotates
and crosses $\phi = \pi$ (mod $2\pi$),
since $M \widehat{\phi}$ changes discontinuously 
by $2\pi$ at that instant.
This is the 2D analog of marking the location of a 1D kink
at the point where $\phi=\pi$ (mod $2 \pi$)
regardless of whether the kink really has a localized structure or not.
This particle-like picture is frequently useful but, by
neglecting spatial distributions, there is a potential loss of
information about the true dynamical state of the system.
On the other hand, the current vorticity $\Omega$ would reveal more 
accurately how localized vortices are.
However, our simulations show
that the topological vorticity $\zeta$ moves together with
a peak of the current vorticity $\Omega$ (Sec.~\ref{sec:numericalvorticity}).
Thus, we use both definitions interchangeably at our convenience.

\newpage
\section{Analysis}
\label{sec:analysis}

In this section we present a perturbative analysis of
the governing equations (\ref{vectorRSJ}),(\ref{goveqn3}). 
Although the analysis is made systematic so that it is
possible to proceed to higher orders,
we show that most of the fundamental features of the row-switched states
can be explained by the leading order of the expansion. 

Before writing down the RS solutions, it is useful
to think of the array with {\em uncoupled} junctions.
This limiting case corresponds to imposing $\psi=0$
 in (\ref{vectorRSJ}), thus reducing the array to a collection of 
uncoupled pendula, independently responding to a constant drive.
The junctions on the horizontal branches~(\ref{goveqn1}), 
whose drive is zero, converge asymptotically to ${\phi^x}^{\ast}=0$ 
due to the damping.
On the other hand, the uncoupled vertical junctions~(\ref{goveqn2}), driven
by $I_{\mbox{\scriptsize dc}}$, have a different dynamical behavior.
For small $\Gamma$, they can converge asymptotically to one of two distinct 
stable states~\cite{steve}:
the superconducting (static) state, which exists only when 
$I_{\mbox{\scriptsize dc}} < 1$, in which the drive is balanced by
the sinusoidal nonlinearity, (i.e.
${\phi^y}^{\ast}=\arcsin{I_{\mbox{\scriptsize dc}}}$); 
or the ohmic (whirling) solution, where 
the first time derivative balances the drive,
and $\phi$ increases at a nearly uniform rate
$\omega = I_{\mbox{\scriptsize dc}}/ \Gamma$
(i.e. the pendulum ``whirls'').
The two attractors may coexist for the same drive,
and hysteresis may occur.

When the junctions are coupled, the simple dynamics of 
the independent junctions is altered, and complex 
spatio-temporal solutions, 
which do not have an analogue in the uncoupled array,
may emerge.
Nevertheless, in the case of the RS solutions
the two states of the driven single junction mentioned above 
(static and whirling) are still valuable ``building blocks'' 
for the analysis of the whole system.
Specifically, the RS states are characterized by alternating regions 
in which the vertical junctions are either stationary (Q regions)
or whirling (S regions).
There are, however, some significant differences
with the uncoupled case.
For instance, the time-averaged current distribution in the coupled array
deviates from the uniform flow.  
Hence, the phases of the stationary junctions can have other values  
than 0 or $\arcsin{I_{\mbox{\scriptsize dc}}}$.
In addition, the rotations of the vertical whirling junctions induce
AC oscillations on the stationary junctions,
and phase-locking among the whirling junctions.
Our analysis in this section is capable of explaining 
most of these effects.

We note that our analysis is restricted to solutions
with  no (static) vortices trapped in any of the $Q$ regions.
The ``no-vortex'' state is expected to 
be most relevant to determine the parameter regime
in which RS states appear.
Similarly, the vertical junctions in the S regions are
assumed to be whirling at almost constant frequency 
neglecting more nonlinear rotation, which is certainly possible, 
as we discuss in Sec.~\ref{sec:conclusion}.

\subsection{Perturbative analysis}
\label{sec:perturbation}

In previous perturbative analyses of junctions and arrays, 
it has been customary to treat $I_{\mbox{\scriptsize dc}}$ 
as a large parameter~\cite{landsberg,WatSwi97}.
However, partially RS states can exist 
only when $I_{\mbox{\scriptsize dc}}$ is sufficiently small,
as we will show below.
Therefore, we use the rotation frequency of the pendulum
$\omega=I_{\mbox{\scriptsize dc}}/\Gamma$ 
as the large parameter in our perturbation.
That is, we will consider the high-frequency limit \cite{likharev}
$\omega \gg 1$,
which can be satisfied for a finite $I_{\mbox{\scriptsize dc}}$ 
if the damping $\Gamma$ is small enough.

Hence, we assume that the variables in the RS states can be 
expanded in powers of $\omega^{-1}$. 
The phases of the horizontal junctions are then approximated by
\begin{equation}
  \phi^x(i,j,t) = \overline{\phi^x_0}(i,j) + 
    \sum_{p=2}^\infty \omega^{-p} \phi^x_p(i,j,\tau) .
\label{expansion1}
\end{equation} \\
while the mesh current is given by
\begin{equation}
  \psi(i,j,t) = \overline{\psi_0}(i,j) + \widetilde{\psi_0}(i,j,\tau)
    + \sum_{p=1}^\infty \omega^{-p} \psi_p(i,j,\tau)
\label{expansion2}
\end{equation}
where we have introduced the normalized time
\[\tau=\omega t= (I_{\mbox{\scriptsize dc}}/\Gamma) \; t .\]
The notation $\overline{(\cdot)}$ expresses time-independent (DC) quantities,
while $\widetilde{(\cdot)}$ are for
the time-dependent (AC) parts whose time average is zero.
Note that the correction of $O(\omega^{-1})$ in~(\ref{expansion1}) 
turns out to be zero, so we neglect that term from start.
The form for the vertical junctions 
must be different in the switched and the quiescent rows.
In the switched rows, 
all the junctions are whirling and
their phases grow, to the lowest order, constantly in time:
\begin{equation}
  \phi^y(i,j,t) = \tau + \overline{\phi^y_0}(i,j)
    + \sum_{p=2}^\infty \omega^{-p} \phi^y_p (i,j,\tau), \;\; j \in S. 
\label{expansion3}
\end{equation}
Meanwhile, in the quiescent rows the junctions are librating and, thus,
the leading order is stationary:
\begin{equation}
  \phi^y(i,j,t) = \overline{\phi^y_0}(i,j) 
    + \sum_{p=2}^\infty \omega^{-p} \phi^y_p (i,j,\tau), \;\; j \in Q .
\label{expansion4}
\end{equation}
We impose $\widetilde{\psi_0}$ and
the higher order terms to be periodic in time.
In general, the period has to be modulated 
and, thus, expanded in $\omega^{-1}$ (strained coordinate).
However, since in the following we will focus on the leading order system,
we set the period to be exactly $2\pi$ in $\tau$ for simplicity.

The perturbative calculation proceeds in the usual way by
substituting (\ref{expansion1})--(\ref{expansion4}) into 
(\ref{vectorRSJ}),(\ref{goveqn3}); 
Taylor-expanding the sine in~(\ref{dimensionalRSJ});
and equating terms of the same order in $\omega$. 
In principle, this procedure can be carried out to higher orders
if secular terms are eliminated by satisfying 
solvability conditions when they arise. 

Balancing the leading order terms, we obtain two sets of equations
since the time-independent (DC) and time-dependent (AC) 
terms must cancel separately. First, the DC terms yield the 
following equations for both types of rows:
\begin{eqnarray}
  \sin \overline{\phi^x_0}(i,j)& =& \overline{\psi_0}(i,j)
    - \overline{\psi_0}(i,j-1)
\label{DCeqn1} \\
  \overline{\phi^x_0}(i,j+1) - \overline{\phi^x_0}(i,j) & = & 2 \pi f
  \nonumber \\ 
    & + & \overline{\phi^y_0}(i+1,j) - \overline{\phi^y_0}(i,j)
\label{DCeqn2}
\end{eqnarray}
and one more equation 
which depends on the type of row (switched or quiescent):
\begin{eqnarray}
  0 &= \overline{\psi_0}(i,j) -\overline{\psi_0}(i-1,j), \;\; j \in S
\label{DCeqn3} \\
   I_{\mbox{\scriptsize dc}} - \sin \overline{\phi^y_0}(i,j) &=
 \overline{\psi_0}(i,j) - \overline{\psi_0}(i-1,j), \;\; j \in Q .
\label{DCeqn4}
\end{eqnarray}
These equations constitute the full DC system.

Similarly, from the AC terms we obtain for both rows
\begin{eqnarray}
  \phi_2^{x''}(i,j,\tau) &=& \widetilde{\psi_0}(i,j,\tau)
    - \widetilde{\psi_0}(i,j-1,\tau)
\label{ACeqn1}\\
  \phi^x_2(i,j+1,\tau) &-& \phi^x_2(i,j,\tau) =
  \nonumber \\
  & &  \phi^y_2(i+1,j,\tau) - \phi^y_2(i,j,\tau) 
\label{ACeqn2}
\end{eqnarray}
where $''$ denotes differentiation twice with respect to  $\tau$.
Moreover, for each type of row we obtain a different equation:
\begin{eqnarray}
  \phi_2^{y''}(i,j,\tau) & = & -\sin (\tau  + \overline{\phi^y_0}(i,j)) 
  \nonumber \\
    & - & \widetilde{\psi_0}(i,j,\tau)
    + \widetilde{\psi_0}(i-1,j,\tau), \;\; j \in S
\label{ACeqn3}
\end{eqnarray}
and
\begin{equation}
  \phi_2^{y''}(i,j,\tau) = 
    - \widetilde{\psi_0}(i,j,\tau) + \widetilde{\psi_0}(i-1,j,\tau), 
    \;\; j \in Q,
 \label{ACeqn4}
\end{equation}
which completes the full AC system.

These systems of equations are to be solved with boundary conditions
\begin{equation}
  \overline{\psi_0} = \widetilde{\psi_0} = 0
\label{DCACBC}
\end{equation}
at the boundary cells.

A simple but important observation can be made at this point.
Using (\ref{DCeqn3}) and (\ref{DCACBC}) at $i=0$ and $N^x+1$ (i.e., 
at the right and left edges), it follows that 
\begin{equation}
  \overline{\psi_0}(i,j) = 0 \quad \mbox{$ \forall i \;$,   if $j \in S$.}
\label{DCswitchedpsi}
\end{equation}
Therefore, the leading order DC mesh current
{\it vanishes in a switched row}~\cite{BCcaveat},
just as it does in the top and bottom boundary cells at $j=0$ and $N^y+1$.
In other words, the switched row is equivalent to having another
boundary row, which splits the array into two.
Thus, to the leading order, a partially row-switched array 
with many switched rows
can be described as a collection of disjoint quiescent regions,
coupled only weakly through the AC component.
This useful picture is exploited later.

The solution of the leading order systems is otherwise non-trivial
since the DC equations (\ref{DCeqn1})--(\ref{DCeqn4}) constitute a
nonlinear algebraic system, 
and the DC solution is in turn needed to
solve the AC system (\ref{ACeqn1})--(\ref{ACeqn4}). 
Thus, in general, they have to be solved numerically --- although 
we show below that useful approximations
can be obtained under certain assumptions.

Once the leading order solutions are found,
the calculation could be carried out to higher orders.
The next order correction leads to 
a particularly simple set of equations:
\begin{eqnarray}
  {\phi_3^x}''(i,j) + \Gamma {\phi_2^x}'(i,j) =
    \psi_1(i,j) - \psi_1(i,j-1) 
\label{nextorderexpansion1} \\
  {\phi_3^y}''(i,j) + \Gamma {\phi_2^y}'(i,j) =
    -\psi_1(i,j) + \psi_1(i-1,j)
\label{nextorderexpansion2} \\
  \phi_3^x(i,j+1) - \phi_3^x(i,j) =
    \phi_3^y(i+1,j) - \phi_3^y(i,j)
\label{nextorderexpansion3}
\end{eqnarray}
for all $\tau$ and
regardless of whether the row $j$ is switched or quiescent.
Terms from the sinusoidal nonlinearity do not come into play
at this order, but further expansions would certainly involve
more complications.

It is important to note, however, that the salient features of 
the solutions observed in the numerics can be explained from the 
leading order equations. 
Therefore, we restrict our analysis to the DC and AC equations 
in the following sections. 
On the other hand, we will also point out 
a remaining problem which is likely to be resolved 
only by considering the higher order corrections.

\subsection{Analysis of the DC equations}
\label{sec:analDCeqn}

The DC equations~(\ref{DCeqn1})--(\ref{DCeqn4}) constitute
a nonlinear algebraic system which must be solved numerically in general. 
However, to gain insight into the system, we will now obtain 
approximate solutions to the system when there is a large asymmetry 
between its two dimensions. 
We will then come back to the full system and discuss its solutions.

\subsubsection{Large aspect-ratio approximation}
\label{sec:largeaspectratio}

Consider the case when all quiescent regions 
in the array are longer horizontally than vertically.
This happens, of course, when the array
itself satisfies $N^x \gg N^y$.
More importantly, 
arrays whose dimensions do not fulfill this condition 
are also broken into smaller, laterally-long,
almost disjoint quiescent regions
after several row-switching events.
Thus, this ``large aspect-ratio'' approximation is important
to characterize the RS states which appear in the course of
the row-switching process.
Remember we also assume that none of the $Q$ regions 
contains static vortices, which could be trapped
for large $N^y$ and $f$, and 
for small $I_{\mbox{\scriptsize dc}}$.
In Section~\ref{sec:existence} we will give an estimate of the values
of $f$ and $N^y$ for which we expect this assumption to be valid.

In such a situation
we expect a nearly ``uniform'' solution in the bulk of the array
with some edge corrections near the right and left boundaries.
Hence, far from the boundaries, we assume the vertical
junctions in the quiescent rows to become {\em independent} of $i$, 
\[ \overline{\phi^y_0}(i,j)=\overline{\phi^y_0}(j) 
\quad \mbox{for $j \in Q$.} \]
On the other hand, we assume a whirling solution~\cite{shinyalong} 
for the switched rows in which
waves with well-defined wavenumbers $k(j)$ propagate:
\begin{equation}
  \overline{\phi^y_0}(i,j) \approx -k(j) i + \delta(j)
  \quad \mbox{for $j \in S$.}
\label{DCassumption1}
\end{equation}
Note that the wavenumber $k(j)$ 
and the phase constant $\delta(j)$ 
may differ from one switched row to another.
The other DC variables
$\overline{\phi^x_0}$, and $\overline{\psi_0}$ are
also assumed to be $i$-independent.

Thus, the DC equations reduce to
\begin{equation}
  \sin \overline{\phi^x_0}(j) = \overline{\psi_0}(j) 
    - \overline{\psi_0}(j-1),
\label{DCsimplified1}
\end{equation}
\begin{eqnarray}
  \overline{\phi^x_0}(j+1) - \overline{\phi^x_0}(j) = 
  2 \pi f - k(j) \quad \mbox{for $j \in S$,}
\label{DCsimplified2}\\
  \overline{\phi^x_0}(j+1) - \overline{\phi^x_0}(j) = 
  2 \pi f \quad \mbox{for $j \in Q$,}
\label{DCsimplified3}
\end{eqnarray}
\begin{equation}
  \overline{\psi_0}(j) = 0 \quad \mbox{for $j \in S$},
\label{DCsimplified4}
\end{equation}
\begin{equation}
  \overline{\phi^y_0}(j) = \arcsin I_{\mbox{\scriptsize dc}} 
  \quad \mbox{for $j \in Q$.}
\label{DCsimplified5}
\end{equation}

This simplified set of equations is still nonlinear but solvable.
We begin by analyzing all the quiescent regions in the array (if any),
delimited by switched regions or by the physical boundaries.
Consider a quiescent region spanning from row $j_1$ to $j_2$
$(\geq j_1)$. 
Such a region contains 
$n=j_2-j_1+2$ rows of horizontal junctions
including the top and bottom borders,
and $n-1$ quiescent rows of vertical junctions.
We emphasize that these vertical phases are all given by (\ref{DCsimplified5}),
thus, $I_{\mbox{\scriptsize dc}}<1$ is necessary for the
existence of partially RS states, where $Q$ rows are present.
From~(\ref{DCsimplified1}) the horizontal phases must satisfy
a telescope sum
\begin{equation}
  \sum_{j=j_1}^{j_2+1} \sin \overline{\phi^x_0}(j)
  = \overline{\psi_0}(j_2+1) - \overline{\psi_0}(j_1-1) =0, 
\label{telescopesum}
\end{equation}
where we have used the fact that both rows $j_2+1$
and $j_1-1$ must be either switched or in the boundary cells, and thus
$\overline{\psi_0}=0$ from (\ref{DCsimplified4}) or (\ref{DCACBC}).
Now, (\ref{DCsimplified3}) can be solved with (\ref{telescopesum}) to
obtain: \cite{roots}
\begin{equation}
  \overline{\phi^x_0}(j+j_1-1) = 2 \pi f \left( j - \frac{n+1}{2} \right)
\label{largeaspectratiophix}
\end{equation}
with $j=1,\ldots,n$. This gives the time-averaged phases
of the horizontal junctions in the bulk of the $Q$ region.
Then, from~(\ref{DCsimplified1}), the mesh current in the same region 
can be computed as
\begin{equation}
  \overline{\psi_0}(j+j_1-1) = \sum_{\ell=1}^{j}
    \sin \overline{\phi^x_0}(\ell)= 
\frac{ \sin (\pi f j)}{\sin (\pi f)} \; \sin \left[ \pi f (j-n) \right ]
\label{psiQ}
\end{equation}
for $j=1,\dots,n-1$.
This procedure allows us to solve for each $Q$ region in the array
independently.

The remaining variables are easy to find.
Recall that $\overline{\psi_0}$ vanishes everywhere 
in the $S$ region.
The rest of the horizontal junctions $\overline{\phi^x_0}$
lie either between two $S$ rows,
or between an $S$ row and a boundary cell.
In either case, it follows from (\ref{DCsimplified1}) that 
\[ \overline{\phi^x_0}(j)=0, {\mbox{\qquad inside a $S$ region}}.\]

Finally, the wavenumbers $k(j)$ for the switched rows ($j \in S$) 
can be calculated from (\ref{DCsimplified2}).
One notices that $k(j)$ can change from a row to another,
depending on the adjacent horizontal junctions $\overline{\phi^x_0}$.
On the other hand, 
if there is an $S$ region with three or more rows,
the wavenumber $k(j)=2\pi f$ for all the rows except for 
the two rows at the top and bottom borders of the region;
this is because $\overline{\phi^x_0}=0$ inside the region.
In this sense, $k_0=2 \pi f$ is the ``natural'' wavenumber 
for $S$ rows.

This concludes the solution of the simplified equations
(\ref{DCsimplified1})--(\ref{DCsimplified5}).
We now exemplify this procedure with four RS states
of an array with $N^y=7$ rows, as depicted in Fig.~\ref{fig:patterns}.
In Section~\ref{sec:numerics} we will compare the predictions with 
our numerics.
\begin{list}{}{
  \setlength{\itemindent}{0mm}
  \setlength{\leftmargin}{2mm}
  \setlength{\labelsep}{1mm}
  \setlength{\labelwidth}{1mm}
}

\item[Pattern 1:] $S = \{1,\dots,7\}$.\\
This is the totally row-switched state in which
there is no $Q$ region.
Thus, the horizontal phases are
\[
  \overline{\phi^x_0} =
  (0, 0, 0, 0, 0, 0, 0, 0).
\]
(The $j$-th component of the vector is $\overline{\phi^x_0}(j)$.
Note this $j$ runs through $1,\dots,8$ for the 7-row array.)
In addition,
$\overline{\psi_0}(j)=0$, and $k(j)=2 \pi f =k_0$
for all rows $j=1,\dots,7$.

\item[Pattern 2:] $S = \{2,4,6\}$, (and so, $Q = \{1,3,5,7\}$).\\
In this symmetric pattern there are four $Q$ regions,
each consisting of only one row, and three one-row $S$ regions.
By solving each $Q$ region independently, we find
\[
  \overline{\phi^x_0} = \pi f
  ( -1,1,-1,1,-1,1,-1,1) .
\]
Then, for the $S$ rows $j=2,4,6$, we have 
$\overline{\psi_0}(j)=-\sin(\pi f)$ 
and $k(j)=4 \pi f$.
That is, the three $S$ rows have an identical wavenumber
but different from the natural $k_0$.

\item[Pattern 3:] $S = \{4\}$.\\
In this case the two symmetric $Q$ regions, rows 1--3
and 5--7, are separated by the central $S$ row.
We obtain:
\[
  \overline{\phi^x_0} = \pi f
  ( -3,-1,1,3,-3,-1,1,3) .
\]
The wavenumber of the $S$ row is $k(4)=8 \pi f$.

\item[Pattern 4:] $S = \{2,3,4,7\}$.\\
In this highly asymmetric switching pattern there are 
two $Q$ regions.
We obtain:
\[
  \overline{\phi^x_0} = \pi f (-1,1,0,0,-2,0,2,0) .
\]
The $S$ rows have the following wavenumbers: 
$k(2)=3 \pi f$, $k(3)=2 \pi f$, and $k(4)=k(7)=4 \pi f$.
Note that the rows 2--4 are contiguous
but all have different wavenumbers.
The row 3 is surrounded by other $S$ rows, 
hence has the natural wavenumber.
Meanwhile, the rows 2 and 4, which are contiguous to
Q regions have different wavenumbers.

\end{list}

A similar bulk approximation can be obtained
for the other limit of the aspect-ratio.
We present this {\it small} aspect-ratio case
in Appendix \ref{sec:smallaspectratio}.

One might wonder what has happened to 
the phase constants $\delta(j)$ of the switched rows~(\ref{DCassumption1}).
Indeed, the $\delta(j)$ have disappeared in the simplified 
system (\ref{DCsimplified1})--(\ref{DCsimplified5}),
making them {\em arbitrary}.
However, simulations show that the switched rows are
weakly coupled,
so that the $\delta$'s drift to some particular values
(if $f \ne 0$).
This phase-locking has been noticed in the completely switched state
and left unexplained \cite{wiesen1,filawiesen,marino}.
As we show in the numerics of Section~\ref{sec:numerics},
it is also a feature of the partially RS states.
The indeterminacy of $\delta$ in our analysis is not merely 
due to the assumption of the whirling 
solution~(\ref{DCassumption1}).
Rather, it is already inherent
in the DC equations (\ref{DCeqn1})--(\ref{DCeqn4})
for which the addition of a constant to all the  
$\overline{\phi^y_0}(i)$ within any switched row
leaves the system unchanged.
Since the drift occurs in a much slower time scale 
than the basic oscillation frequencies \cite{shidelta},
we conjecture that the $\delta(j)$ could be determined from 
solvability (or secularity) conditions that might arise
from higher orders of the  expansion.
That was the case in one-dimensional series arrays~\cite{WatSwi97} where 
a similar slow phase drift and eventual locking was explained in that manner.
However, it is beyond the scope of this paper 
to develop a similar calculation for the 2D array,
and, in the following, we will use the values of $\delta(j)$ 
obtained from the numerical simulations.

\subsubsection{Solving the full DC equations}
\label{sec:analfullDCeqn}

We now consider how to solve the full DC system
beyond the bulk approximation---a problem which requires, in general,
numerical solution.
It is important, however, to note that the decoupling of the equations
introduced by the switched rows is still present
so that the problem reduces to calculating {\em static}
solutions of smaller arrays.

The important point to recall is given by~(\ref{DCswitchedpsi}): 
the mesh current is still zero in all $S$ rows.
This breaks the array into disjoint $Q$ regions,
as far as the leading order DC part is concerned.
Mathematically, this means that
equations (\ref{DCeqn1}),(\ref{DCeqn2}),
and (\ref{DCeqn4})
are closed within each $Q$ region, and can be solved independently.
This system is identical to the superconducting (static) equations
for an isolated 2D array of the same size as the $Q$ region.
When this sub-problem of finding the static solutions for the independent
Q regions is solved,
the remaining unknowns, $\overline{\phi^y_0}$ in the $S$ regions,
can be determined from (\ref{DCeqn2}).
This two-step procedure is completely analogous to 
the one used in the large aspect-ratio approximation,
except that $\overline{\phi^y_0}$
in the $Q$ rows now depends on $i$, and, thus, 
$\overline{\phi^y_0}(i,j)$ in the $S$ rows cannot have the form
given in~(\ref{DCassumption1}).

How do we obtain the static configurations?
Since a $Q$ region can take any size in the $j$-direction (up to $N^y$),
we need, in short, a general calculation scheme
of static states for an arbitrary rectangular array.
An analytical formula is not known
even for the no-vortex solutions 
(one of the many possible superconducting states)
we are primarily concerned with.
Thus, they must be found numerically \cite{teitel,2Dsc}.
A rare exception is the {\it ladder array}, of size $N^x \times 1$,
for which an accurate analytical approximation 
has been obtained~\cite{ladderdep}.
It shows
that the full static solution differs from the bulk approximation
in the existence of skin layers near the left and right edges.
Crucially, resolving the phases in the skin layers is central to  
the existence and stability of the static solution.
The ladder case is special but important
since it is the most persistent 
in the parameter space among $Q$ regions of a given width~\cite{stroud,2Dsc}.
In Section~\ref{sec:existence} we will connect the
stability of the RS patterns with the stability of
the static states.

\subsection{Analysis of the AC equations}
\label{sec:analACeqn}

We now study the AC system~(\ref{ACeqn1})--(\ref{ACeqn4}).
We only need to note that this is a {\em linear} system which is
forced by the sinusoidal drive
$\sin (\tau + \overline{\phi^y_0})$.
Therefore, if the DC solution is known, 
the AC system is simple to analyze.

Assuming that the homogeneous part 
simply decays, the solution locks to the forcing
and the time-dependence can be factored out as:
\begin{equation}
  \left[ \begin{array}{c}
     \phi^x_2 \\ \phi^y_2 \\ \widetilde{\psi_0}
  \end{array} \right] (i,j,\tau) = \left[ \begin{array}{c}
     A \\ B \\ C
  \end{array} \right] (i,j) \exp (\tau \sqrt{-1}) + \mbox{c.c.}
\label{ACsoln}
\end{equation}
where ``c.c.''\ denotes complex conjugate. Then,
the spatially-dependent complex amplitudes must satisfy
\begin{eqnarray}
  -A(i,j) = C(i,j)-C(i,j-1)
\label{ACcoefeqn1} \\
  -B(i,j) = -C(i,j)+C(i-1,j)+f(i,j)
\label{ACcoefeqn2} \\
  A(i,j+1)-A(i,j) = B(i+1,j)-B(i,j)
\label{ACcoefeqn3}
\end{eqnarray}
with
\begin{equation}
  f(i,j) = \left\{ \begin{array}{cl}
    \frac{\sqrt{-1}}{2} \exp \left(
      \overline{\phi^y_0}(i,j) \sqrt{-1} \right) &
      \mbox{if $j \in S$} \\
    0 & \mbox{if $j \in Q$.}
  \end{array} \right.
\label{ACcoefeqn4}
\end{equation}

Eliminating $A$ and $B$ from the equations,
we obtain a discrete Poisson equation for $C$:
\begin{eqnarray}
  \Delta C = -\mu
\label{discreteLaplace}
\end{eqnarray}
with the source term
\begin{equation}
  \mu(i,j) = f(i,j)-f(i+1,j) 
\label{sourceterm}
\end{equation}
and, from~(\ref{DCACBC}), boundary conditions  
\begin{equation}
  C=0 \quad \mbox{in the boundary cells.}
\label{ACcoefBC}
\end{equation}

In the rectangular domain this problem can be solved
via the double discrete Fourier-sine series
\begin{equation}
  C(i,j) = \sum_{m=1}^{N^x} \sum_{n=1}^{N^y}
  \widehat{C}_{m,n}
  \sin \left( \frac{mi\pi}{N^x+1} \right)
  \sin \left( \frac{nj\pi}{N^y+1} \right)
\label{Fouriersine}
\end{equation}
with
\begin{equation}
  \widehat{C}_{m,n} = 
  \frac{1}{a^2_{m,n}}
  \sum_{i=1}^{N^x} \sum_{j=1}^{N^y} \mu(i,j)
  \sin \left( \frac{im\pi}{N^x+1} \right)
  \sin \left( \frac{jn\pi}{N^y+1} \right)
\label{invFouriersine}
\end{equation}
where
\begin{eqnarray}
  a^2_{m,n} & = & (N^x+1)(N^y+1) \nonumber \\
  & & \left\{
  \sin^2 \left( \frac{m\pi}{2(N^x+1)} \right)
  + \sin^2 \left( \frac{n\pi}{2(N^y+1)} \right)
  \right\} .
\label{Fourierfactor}
\end{eqnarray}
Finally, $A$ and $B$ are determined from 
(\ref{ACcoefeqn1}) and (\ref{ACcoefeqn2}).
This completes the analysis of the leading-order equations.

\newpage
\section{Numerics}
\label{sec:numerics}

\subsection{Finding RS states in simulations}

To test the validity of the analysis
developed in the previous section we now compare its predictions 
with numerical results.
The full governing equations~(\ref{vectorRSJ}) and (\ref{goveqn3}),
together with the boundary conditions~(\ref{BC}), are integrated using 
the standard fourth and fifth order Runge-Kutta integrator with  
adaptive time step.
Ours is an elementary non-optimized version of the previous mesh-formulated 
code~\cite{joelrs,ETthesis} which 
enables us to switch between no-inductance and simple 
inductance models. The results presented here are all obtained 
neglecting inductances.

As stated above, we also neglect the effects of 
temperature and disorder.
Since most of the analysis have assumed 
the large aspect-ratio approximation,
we study an array with $N^x=31$ and $N^y=7$, 
with small damping $\Gamma=0.2$ and
a moderate external field $f=0.1$.
We use as initial conditions
the predicted large aspect-ratio DC
approximations $\overline{\phi_0^{x,y}}$
(and the corresponding first time derivatives).
They are expected to be close enough to the true RS states 
to facilitate convergence,
but we leave the AC part to be adjusted by the system.
We choose a value for $I_{\mbox{\scriptsize dc}}$ between 0 and 1,
and monitor whether the ensuing dynamical state is
indeed the attempted RS pattern.

The system, of course, does not always converge to 
the row-switched state we have targeted;
the chosen initial condition may be out of the basin
of attraction of the target state,
or the state may not exist,
or it may be unstable for the chosen parameters.
The outcome from using ``wrong'' parameters is,
as far as we have tested, as follows:
If $I_{\mbox{\scriptsize dc}}$ is too large,
then vortices start to enter in some of the rows
we have initially set quiescent;
if $I_{\mbox{\scriptsize dc}}$ is too small,
then the rows we have set switched cannot maintain 
the whirling motion,
and exhibit retrapping, 
become quasi-periodic,
or show highly nonlinear oscillations.
In those cases, we then adjust $I_{\mbox{\scriptsize dc}}$ until we find 
the clean periodic RS solutions which we aimed at.

Not only must we tune $I_{\mbox{\scriptsize dc}}$, but the damping 
parameter $\Gamma$ must be small enough in order to find clean RS states.
If $\Gamma$ is too large, it is difficult to find any partially 
RS states at all. 
For intermediate values, such as $\Gamma=0.4$,
some RS patterns are observed, but some others cannot be found.
For the underdamped case $\Gamma=0.2$ studied here,
it becomes easy to find an appropriate range of
$I_{\mbox{\scriptsize dc}}$ in which the system converges
to the expected RS pattern.
This dependence on the damping is in qualitative agreement 
with experimental findings~\cite{experimrs}.
It is also consistent with our assumption of the high-frequency limit
since a smaller $\Gamma$ for a given $I_{\mbox{\scriptsize dc}}$
corresponds to a larger $\omega$.

Generally, patterns with large quiescent regions 
are more difficult to obtain; 
for example, the RS state $S=\{ 1 \}$
(with one $Q$ region of 6 rows)
has a smaller interval of suitable $I_{\mbox{\scriptsize dc}}$
than the symmetric Pattern~3 
$S=\{ 4 \}$ (with two $Q$ regions of 3 rows),
even though both states have only one $S$ row.
These observations and the above parameter dependences will
be discussed in more detail in Sec.~\ref{sec:existence}.

\begin{figure}[tbp]
\centerline{\psfig{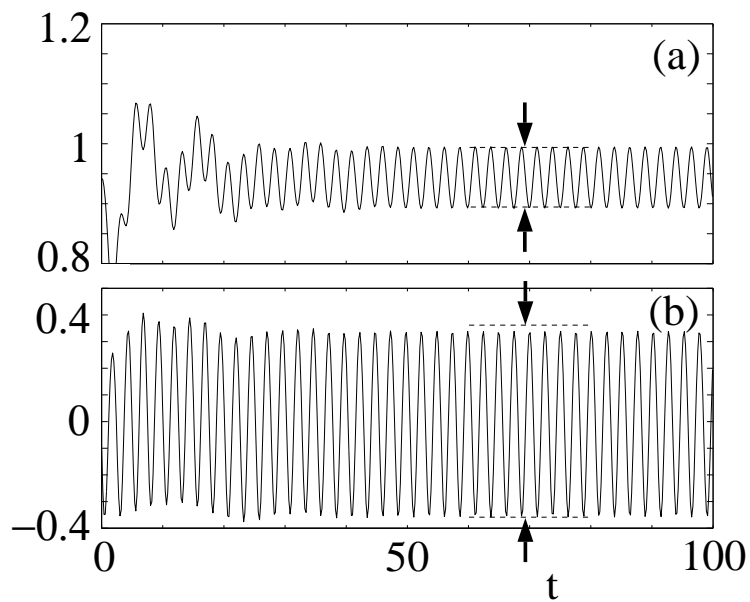}}
\vspace{5mm}
\caption{
Time evolution in the central (switched) row of Pattern 3:
(a) horizontal phase $\phi^x(16,4,t)$
and (b) mesh current $\psi(16,4,t)$.
After a short transient, 
the solution converges onto a periodic attractor
with DC values and AC amplitudes 
(shown together as the bands delimited by the dotted lines)
well predicted by the analytical formulas 
in Sec.~\protect{\ref{sec:analysis}}.
}
\label{fig:convergence}
\end{figure}

Before presenting detailed comparisons between numerics
and analysis for Patterns 1--4, we first
illustrate convergence in Fig.~\ref{fig:convergence}. There we show 
the time evolution of two variables in the array for Pattern 3,
using $I_{\mbox{\scriptsize dc}}=0.5$.
Since the initial condition (taken as the bulk approximation) 
is not a solution of the {\it full} system, there is a short 
transient ($t < 50$) until the system settles onto a periodic attractor.
Recall that only row 4 is switched in this pattern.
Figure \ref{fig:convergence}(a) shows 
the phase $\phi^x(16,4)$ of a horizontal junction
adjacent to the switched row and in the middle of the row,
where the large aspect-ratio (bulk) approximation is
expected to be valid.
The approximated average value is 
$3 \pi f \approx 0.94$, as predicted in Sec.~\ref{sec:analysis}.
Similarly, the mesh current in the central cell $\psi(16,4)$, 
shown in Fig.~\ref{fig:convergence}(b),
is $\psi=0$ on average with some oscillations,
as expected in any switched row.
Not only the average values but the AC amplitudes 
are also well estimated from the AC leading order equations,
as demonstrated in Fig.~\ref{fig:convergence}.

\subsection{Vortex motion}
\label{sec:numericalvorticity}
 
\begin{figure}[tbp]
\centerline{\psfig{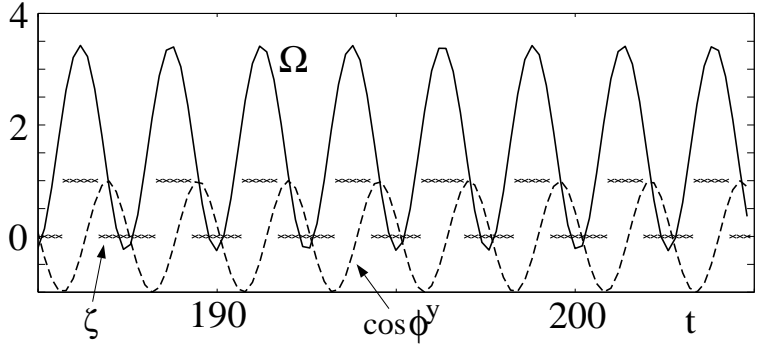}}
\vspace{5mm}
\caption{
Time evolution of the vorticities 
in the middle of the array of Pattern 3.
The solid curve depicts the current vorticity $\Omega(16,4,t)$,
while the topological vorticity $\zeta(16,4,t)$
switches discontinuously between 0 (no vortex) 
and 1 (one vortex in the cell).
This discontinuous ``tagging'' of the position of the vortex is clarified by
the dotted curve, which corresponds to $\cos \phi^y(16,4,t)$. 
Inspection of that magnitude indicates that everytime it becomes $-1$ 
(i.e., the phase is equal to $\pi$)
one topological vortex enters the cell 
(and $\zeta$ is increased by one).
}
\label{fig:vortexconvergence}
\end{figure}

We illustrate now the two vorticities defined in 
Section~\ref{sec:formulation}.
In Fig.~\ref{fig:vortexconvergence} we show
the current vorticity $\Omega$
and the topological vorticity $\zeta$ 
both at the central cell (15,4) of Pattern 3 after convergence.
They display similar periodic behavior 
though $\Omega$ is continuous whereas $\zeta$ switches
discontinuously between 0 and 1.
$\zeta$ becomes unity
when a charge enters the cell,
which occurs in this case when $\phi^y(16,4)$,
the left junction, crosses $\pi$ (modulo $2 \pi$). Therefore,
$\zeta$ becomes unity when $\cos \phi^y(16,4)=-1$, 
as shown in the figure.
Similarly, when the right junction $\phi^y(17,4)$ (not shown) turns
and crosses $\pi$,
the charge $\zeta$ is reset to zero.

\begin{figure}[tbp]
\centerline{\psfig{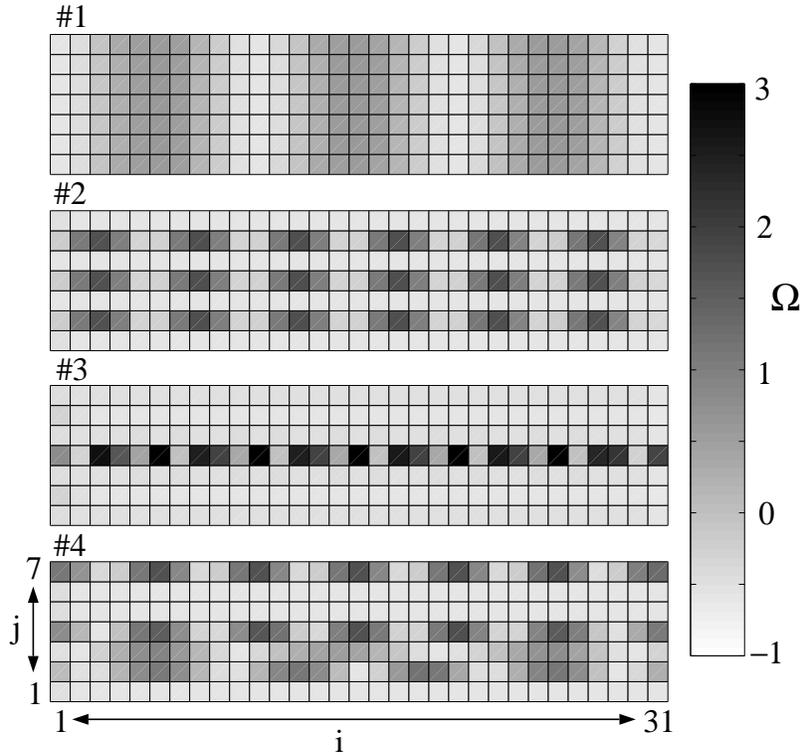}}
\vspace{5mm}
\caption{
Snapshots of Patterns 1--4 showing 
spatial distributions of the current vorticity $\Omega$
as density plots.
Dark regions correspond to large positive $\Omega$.
Compare them with Fig.~\protect{\ref{fig:patterns}}
where the same spatial patterns are shown
in terms of the topological charge $\zeta$.
We observe that the topological vortices
are generally located on peaks of $\Omega$,
and propagate locked to the underlying wave.
}
\label{fig:2Dvortex}
\end{figure}

As a complement to the time evolution of the vorticities in one cell,
we now show snapshots of their spatial distributions for 
all Patterns 1--4 in Fig.~\ref{fig:2Dvortex}.
Each cell is shaded according to the value of
the current vorticity $\Omega(i,j)$:
dark regions indicate positive large $\Omega$,
while bright parts correspond to negative $\Omega$.
The same snapshots, but showing the topological charges $\zeta$,
are given in Fig.~\ref{fig:patterns}.
Even though $\Omega$ represents the spatial structure more clearly,
we observe that a charge in a cell 
corresponds to a peak of $\Omega$,
and that the charges propagate through the array
on top of the underlying wave.

\begin{figure}[tbp]
\centerline{\psfig{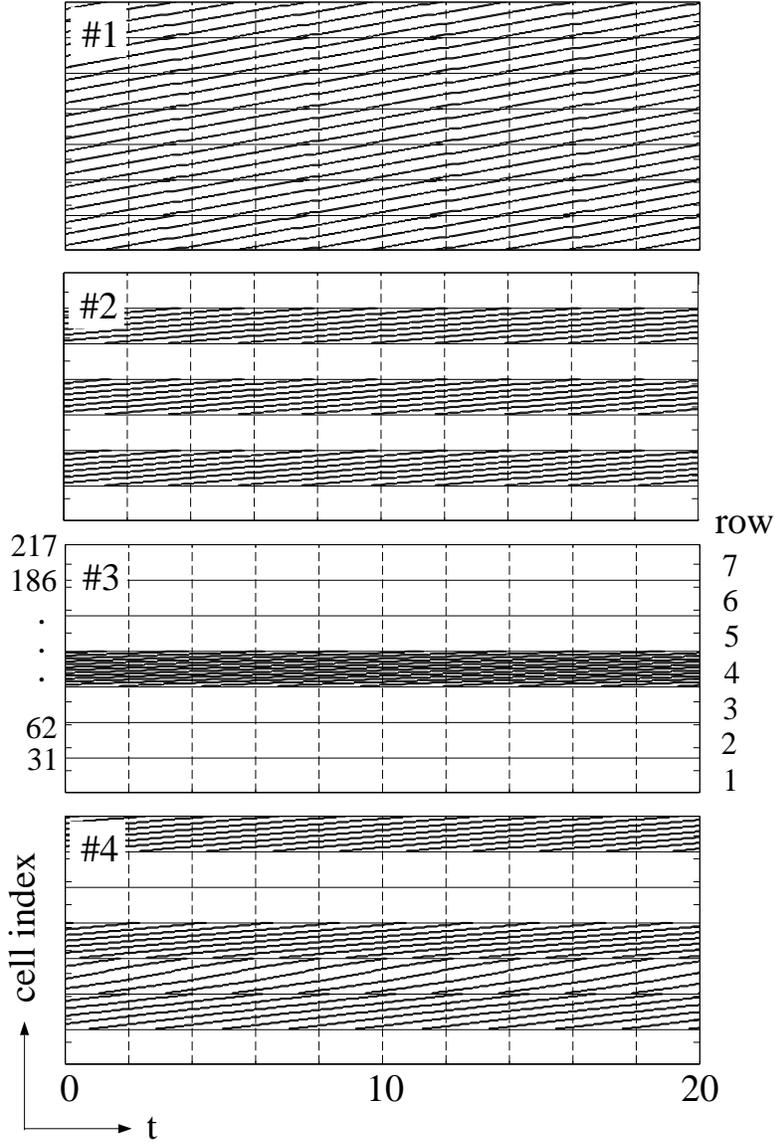}}
\vspace{5mm}
\caption{Space-time plots of the propagation of topological vortices
for Patterns 1--4.
The vertical (space) axis is the cell index: the cell $(i,j)$
is indexed one-dimensionally by $i+N^y (j-1)$ by juxtaposing row after row.
Within each of the symmetric Patterns (1--3),
the vortices in the switched rows have the same wavelength
and are in-phase.
However, in the asymmetric Pattern 4, the spatial wavenumbers 
differ from row to row.
}
\label{fig:vortexpath}
\end{figure}

Thus, we can use $\zeta$
to visualize the wavelength and the propagation speed in each row.
In all Patterns 1--4,
the charges move across the array
at a nearly constant speed,
as seen in the space-time plots 
of $\zeta$ in Fig.~\ref{fig:vortexpath}.
They propagate only through the S rows,
and are apparently in-phase in all rows for Patterns 1 and 2.
However, in Pattern 4 
the S rows are not in-phase, and
the propagation velocities vary from row to row.
Thus, the simplistic picture that vortices carry all the flux and 
move with the same speed in all the S rows within a pattern~\cite{joelrs}
leads to estimated speeds in disagreement with our simulations.
This further proves that the underlying assumption that the 
topological vortices
are particle-like objects which concentrate the flux is not accurate.
Instead, the RS solutions are not localized states and the flux is spatially 
distributed, as suggested by previous work~\cite{geigenlobb,marino}
and demonstrated in our analysis.
Therefore, in these states, the topological vortices merely mark 
where the rotating junctions
cross $\pi$ (mod $2\pi$) (see Sec.~\ref{sec:vorticities}),
and they travel at the phase velocities of the underlying 
(non-localized) waves.
Our analysis correctly estimates
the spatial wavenumbers (thus, the propagation speeds)
as shown below.

\subsection{Spatial structures after convergence}

We now present a quantitatively comparison of the analysis
of Sec.~\ref{sec:analysis} to numerical simulations.
The analytical predictions correspond to the large aspect-ratio 
(bulk) approximation both for the DC and the AC components.
For the numerics, we simulate a $31 \times 7$ array,
and the system is allowed to converge to periodic solutions 
for Patterns 1--4 using $\Gamma=0.2$, $f=0.1$ and 
$I_{\mbox{\scriptsize dc}}=0.6$
(except for Pattern 3, in which $I_{\mbox{\scriptsize dc}}=0.5$
had to be used).

\begin{figure}[tbp]
\centerline{\psfig{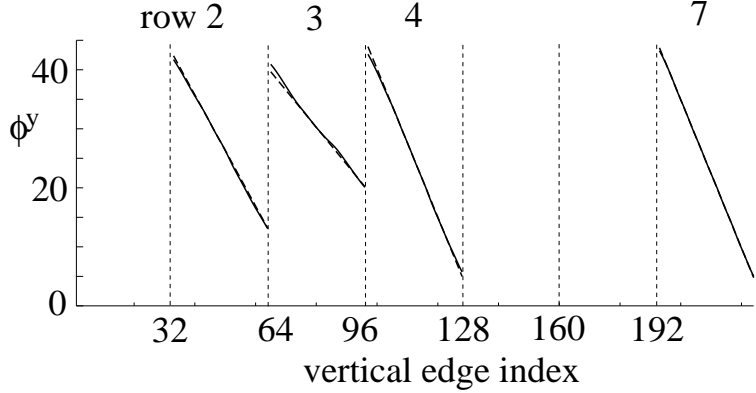}}
\vspace{5mm}
\caption{
A snapshot of the vertical phases $\phi^y$ in the switched rows 
$S=\{2,3,4,7\}$ of Pattern 4
in a $N^x=31$ by $N^y=7$ array.
Each solid line connects the numerical phases of the 32 junctions 
in each switched row.
The dashed lines (almost overlapping with the solid ones) 
are the analytical approximation 
in Section~\protect{\ref{sec:analysis}}
which predict the observed spatial wavenumber very well.
The horizontal axis denotes the ``vertical edge index,''
which numbers the vertical junctions consecutively as 
$i+(N^x+1) (j-1)$ for $i=1,\dots,N^x+1$ 
and $j=1,\dots,N^y$.
This enables us to display the 2D array
in a single axis by juxtaposing one row after the other.
As a guide to the eye, vertical dotted lines are added
to separate the rows.
}
\label{fig:wavenumber}
\end{figure}

We first check the predicted spatial wavenumbers $k(j)$
in the $S$ rows just discussed above.
In Fig.~\ref{fig:wavenumber} we show a ``snapshot''
of the $\phi^y$ in the $S$ rows (2,3,4,7)
of the non-trivial Pattern 4. To ease the display and comparison
of the numerical results, we have juxtaposed the rows one after the other.
Within each row, the spatial dependence is clearly linear,
thus justifying the whirling mode assumption~(\ref{DCassumption1}). 
The predicted wavenumbers 
$k_2=3 \pi f$, $k_3=2 \pi f$, and $k_4=k_7=4 \pi f$ (dashed lines) 
are almost indistinguishable from the numerics (solid lines)
except for small deviations close to the edges.

Recall that in our analysis of the DC equations
 the inter-row phase differences 
$\delta(j)$ are predicted to be arbitrary in~(\ref{DCassumption1}). 
Hence, only the {\it slope} of the spatial dependence is known and 
the dashed lines are adjusted to match at the center of each row.
Conversely, this is a way to determine the $\delta(j)$ 
from the numerical simulations.
For the four Patterns, we obtain: 
\begin{list}{}{
  \setlength{\itemindent}{0mm}
  \setlength{\leftmargin}{2mm}
  \setlength{\labelsep}{1mm}
  \setlength{\labelwidth}{1mm}
}

\item[Pattern 1:] $\delta(1)=\delta(7)=0$, $\delta(2)=\delta(6)=0.05$, \\
$\delta(3)=\delta(4)=\delta(5)=0.1$.

\item[Pattern 2:] $\delta(2)=\delta(4)=\delta(6)=0$.

\item[Pattern 3:] $\delta(4)=0$.

\item[Pattern 4:] $\delta(2)=-1.8$, $\delta(3)=-4.7$, 
$\delta(4)=0.2$,\\ $\delta(7)=0$. 

\end{list}
Note that in each case one $\delta(j)$ is set to zero
and taken as the reference, which
is equivalent to choosing the origin of $t$.
In the following, we will use these numerical values of $\delta$ when
needed (most importantly, for the analytical values of the AC components). 

\begin{figure}[tbp]
\centerline{\psfig{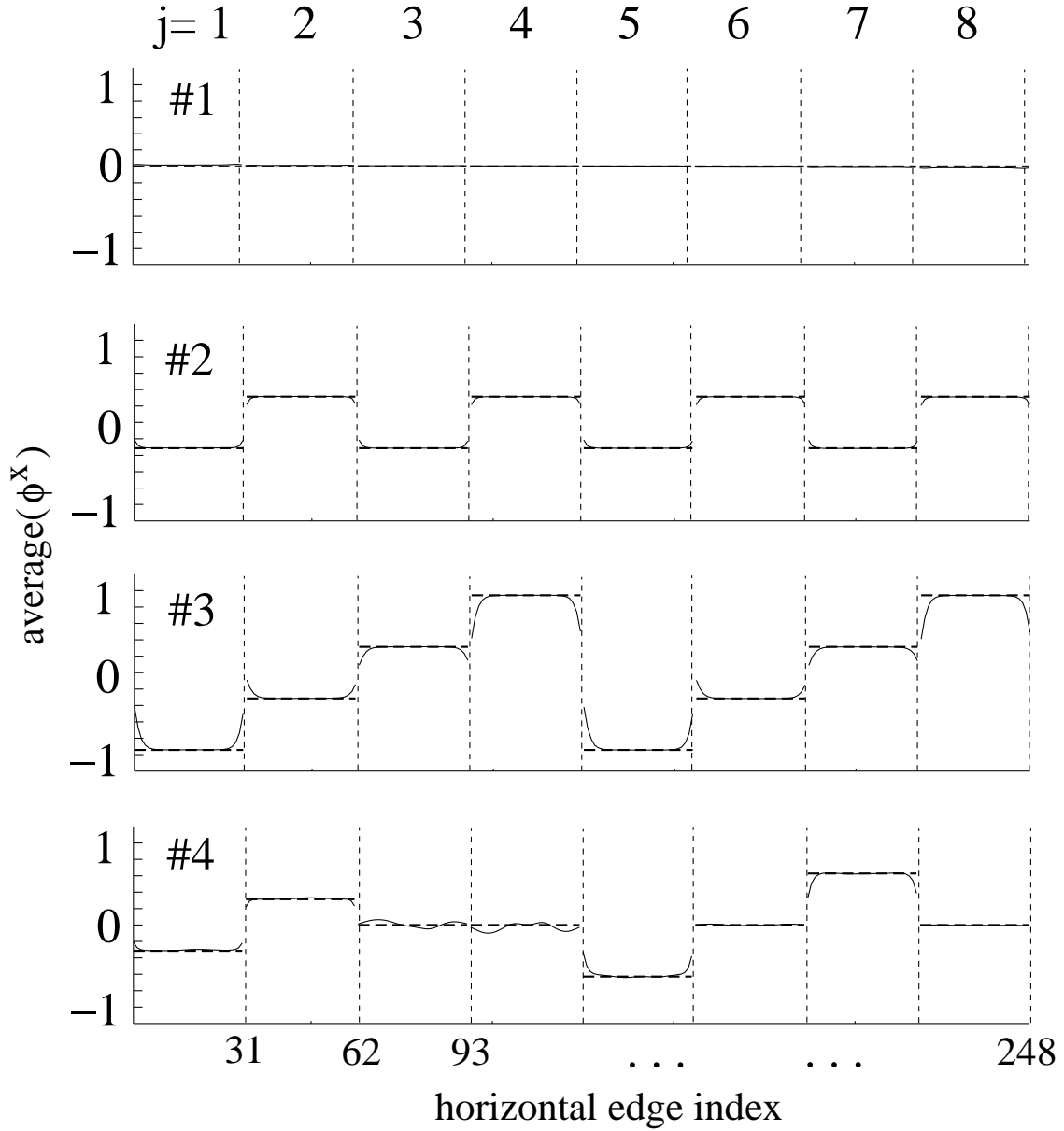}}
\vspace{5mm}
\caption{
The DC values of $\phi^x$ for Patterns 1--4,
showing the spatial distribution of the average horizontal phases.
The horizontal axis is the ``horizontal edge index,''
defined as $i+N^x (j-1)$ for $\phi^x(i,j)$.
There are $N^y+1=8$ horizontal edges
so that $j$ runs from 1 to 8.
For each $j$, the $N^x=31$ phases in the same row are connected.
The dotted lines are from the large aspect-ratio approximation
which accurately estimates the numerical results in the bulk of the array.
The DC values are predicted to be multiples of $\pi f$.
The approximation neglects the effects of the left and right edges,
and, thus, inevitably misses the skin layers at both lateral boundaries. 
Vertical dashed lines mark the separation between $j$'s.
}
\label{fig:hjcnave}
\end{figure}

\begin{figure}[tbp]
\centerline{\psfig{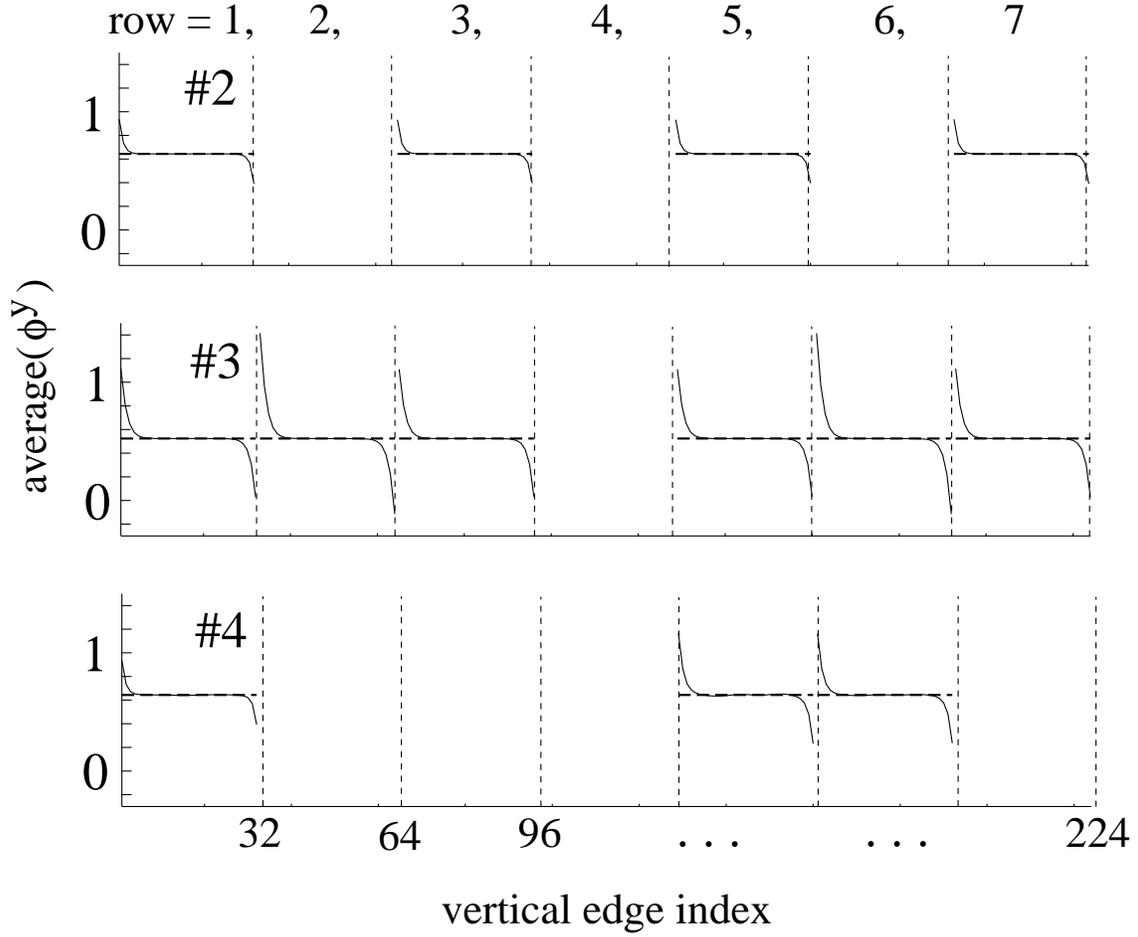}}
\vspace{5mm}
\caption{
The DC values of $\phi^y$ for Patterns 2--4.
Switched junctions are unbounded, thus not shown.
Therefore, Pattern 1 is absent in this figure 
since all the junctions are switched.
The horizontal axis is the vertical edge index, defined 
in Figure~\protect{\ref{fig:wavenumber}}.
For each quiescent row its $N^x+1=32$ phases are connected.
The large aspect-ratio approximation,
dashed lines at $\overline{\phi^y} = \arcsin I_{\mbox{\scriptsize dc}}$,
is a good approximation in the bulk,
but misses the lateral skin layers.
}
\label{fig:vjcnave}
\end{figure}

Next, we compare the predicted DC values with the numerical mean values 
after convergence in Figures~\ref{fig:hjcnave} and~~\ref{fig:vjcnave}.
As we showed in Fig.~\ref{fig:convergence}(a),
each horizontal junction $\phi^x$ librates around some DC value 
after convergence.
These average values are plotted (solid lines) and compared to 
the large aspect-ratio approximation (dotted lines) in 
Fig.~\ref{fig:hjcnave}.
The prediction is uniform within each row because edge effects
were neglected---consequently, it works well everywhere except close to the
right and left ends.
In Fig.~\ref{fig:vjcnave}
we show the DC values of the vertical junctions 
$\phi^y(i,j)$.
Note that the phases of whirling junctions are unbounded, thus not shown.
In particular, Pattern 1 is left out from this figure since 
all $\phi^y$ are switched.
Again, in the bulk of the array the approximation
($\phi^y=\arcsin I_{\mbox{\scriptsize dc}}$) holds, 
but near the edges there is a significant deviation.

\begin{figure}[tbp]
\centerline{\psfig{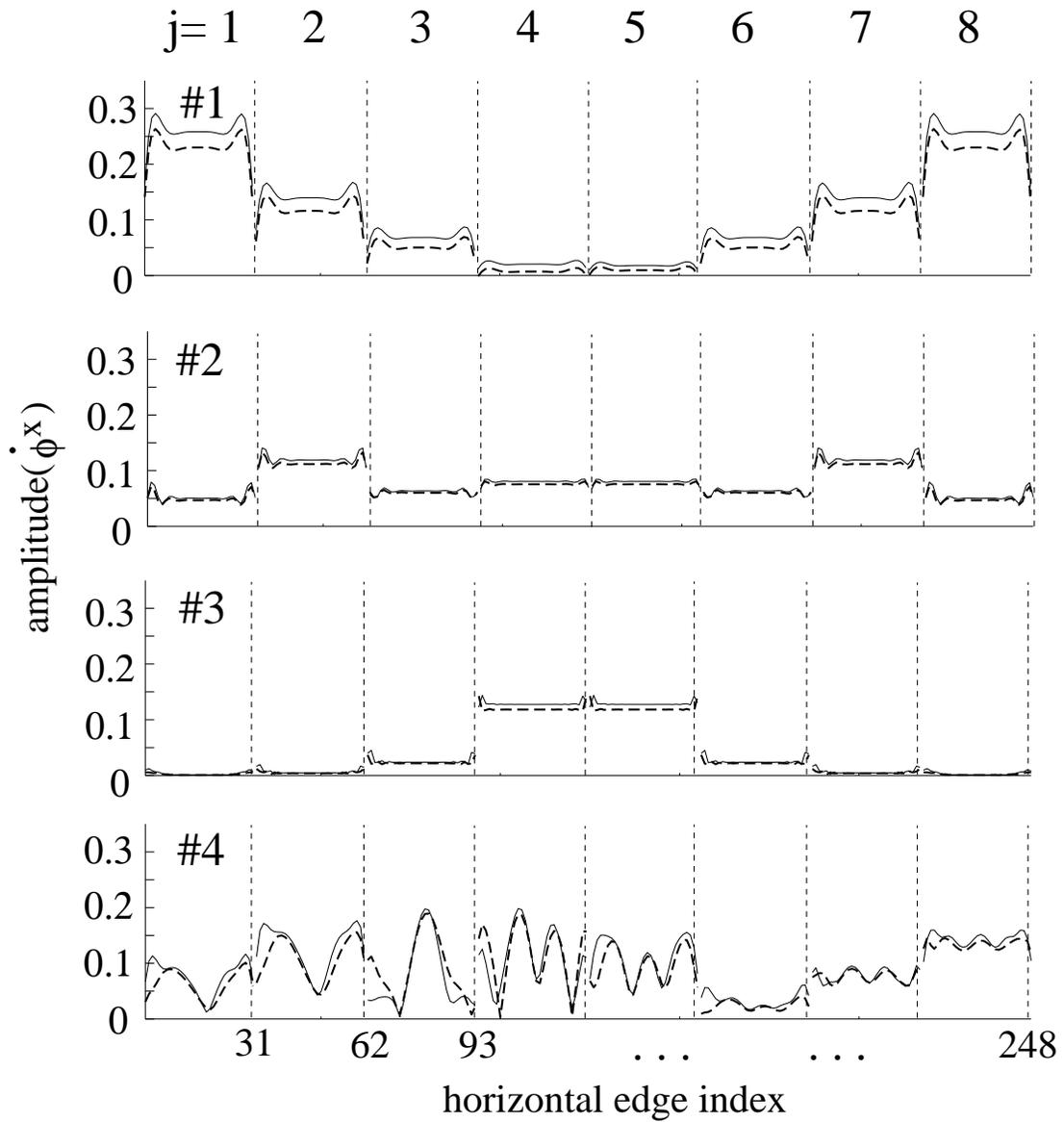}}
\vspace{5mm}
\caption{
Dimensionless AC voltage amplitudes $\dot{\phi^x}$ for Patterns 1--4,
plotted versus the horizontal edge index.
The large aspect-ratio approximation is shown as dashed curves.
There are some quantitative discrepancies,
but the approximation captures the spatial distribution 
of the amplitudes.
}
\label{fig:hjcnamp}
\end{figure}

\begin{figure}[tbp]
\centerline{\psfig{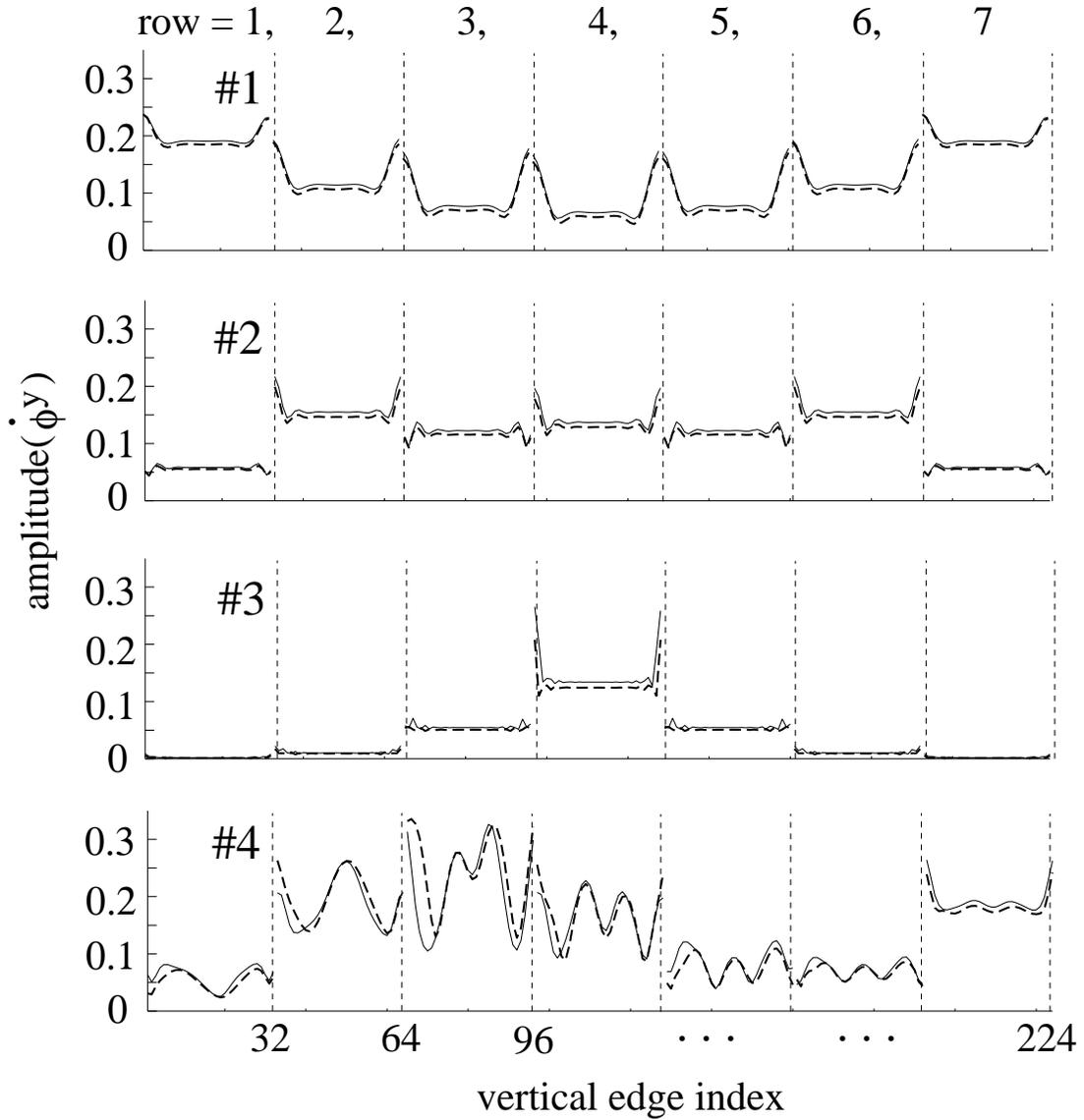}}
\vspace{5mm}
\caption{
Dimensionless AC voltage amplitudes $\dot{\phi^y}$ for Patterns 1--4,
plotted against the vertical edge index.
Again, the large aspect-ratio approximation, shown as dashed curves,
can describe the spatial distribution fairly well.
}
\label{fig:vjcnamp}
\end{figure}

In a similar manner, we show 
in Figs.~\ref{fig:hjcnamp} and \ref{fig:vjcnamp} 
the AC amplitudes of the horizontal and
vertical junctions, respectively;
that is, the $|A(i,j)|$ and $|B(i,j)|$ 
calculated in Section~\ref{sec:analACeqn}.
As seen in both figures,
symmetric Patterns 1--3 have rather constant amplitudes
throughout each row, except near the left and right edges.
The asymmetric Pattern 4 shows spatial fluctuations.
Our estimates, shown as dotted lines,
reproduce the spatial structure fairly well.
It is quite remarkable that our approximation roughly 
captures the behavior 
at the right and left boundaries, when 
we have used the bulk approximation 
$\overline{\phi^y_0}$ (together with the numerical $\delta(j)$)
to solve the AC system.

\begin{figure}[tbp]
\centerline{\psfig{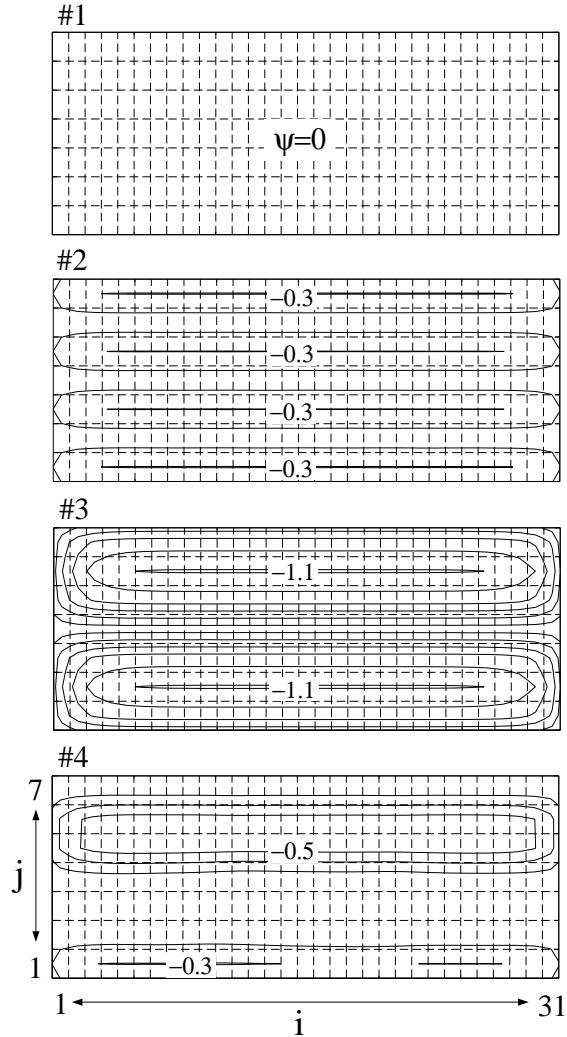}}
\vspace{5mm}
\caption{
Level curves of the DC mesh current $\psi$ for 
Patterns 1--4,
indicating how the induced circulating currents flow.
The total current flow is the superposition of the induced
flow and the injected uniform current flow.
Contour levels at $-0.1,-0.3,\dots,-1.1$ are drawn on the 2D grid
of the $N^x=31$ by $N^y=7$ array.
Pattern 1 shows little deviation from the uniform current flow
on average, thus $\psi =0$ and no curves appear.
In the other patterns, 
the DC values of $\psi$ in the switched rows are zero,
while the values are negative in the quiescent rows.
Therefore, currents circulate in the clockwise direction
in each quiescent region ``along'' the level curves shown.
Strictly, the currents are restricted to the grid, 
but the level curves provide an intuitive description of the flow.
Note that the boundary condition $\psi=0$ is imposed
at a half cell outside of the array borders;
this explains why some of the contour curves intersect
the array edges.
}
\label{fig:DCpsicontours}
\end{figure}

Since the mesh current is 
determined from the phase configurations $\phi^{x,y}$,
it also compares well with the large aspect-ratio approximation.
Thus, we do not display the quantitative comparison of $\psi$,
and instead present more descriptive 2D contour plots 
of the {\it numerical} $\psi$ on the $31 \times 7$ array geometry.
The contour curves of the DC component of $\psi$ are
shown in Fig.~\ref{fig:DCpsicontours}.
If the 2D array were continuous, 
the induced currents would flow along these curves  on average.
Since the array is discrete, the flow is restricted to the branches,
but the level curves still describe roughly
the way the currents circulate.
Furthermore, the DC values in the $S$ rows are nearly zero, as expected.
In the $Q$ regions, currents circulate in the clock-wise direction
($\overline{\psi_0} < 0$) on average.
This would induce a magnetic field through the $Q$ regions
in the opposite direction to the external field $f$.
Although it is interesting to ask 
whether the induced field cancels the external one to produce a Meissner-like
region, that question only makes sense when all inductances are included.
Note also that all contour plots are almost left-right symmetric,
but show a slight asymmetry.
This is presumably due to the presence of edges
and the preferred direction (${\bf \hat x}$) of propagation of the waves 
across the $S$ rows.
Such details are not captured by the bulk approximation
and the full solution of the DC equations becomes necessary.

\begin{figure}[tbp]
\centerline{\psfig{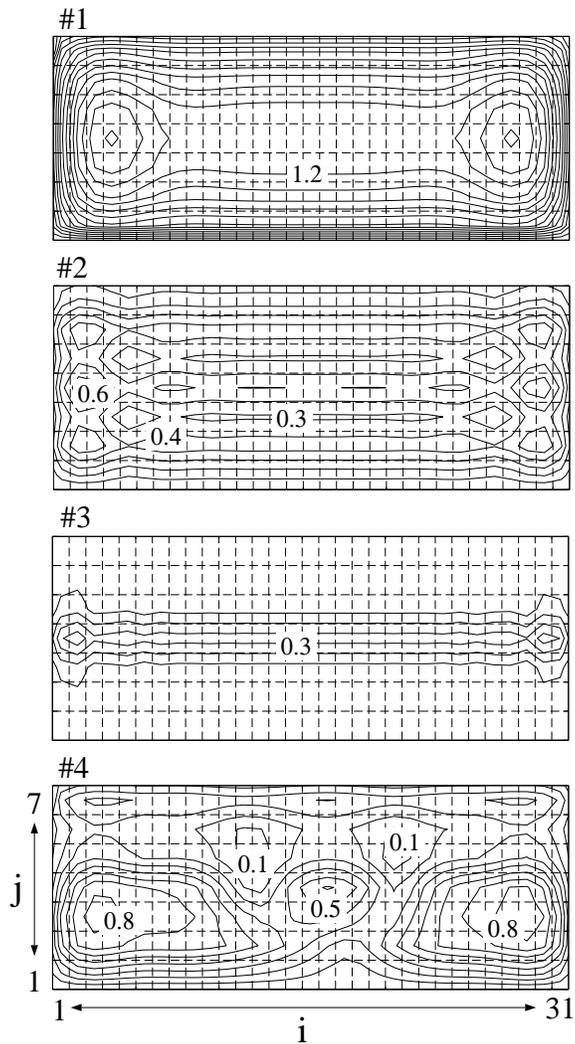}}
\vspace{5mm}
\caption{
Level curves of the AC amplitudes of the mesh current $\psi$
for Patterns 1 (top) to 4 (bottom),
on the 2D grid of $31 \times 7$ cells.
Contour levels at $0.1,0.2,\dots,1.5$ are shown.
The magnitudes are generally large in the switched rows,
but even quiescent rows have some oscillations
and, thus, are not purely superconducting.
Our leading-order analysis in Sec.~\protect{\ref{sec:analACeqn}}
predicts that these AC oscillations obey
the discrete Poisson equation 
with forcing originating from the DC components.
The figure shows nodal structures typical in solutions to such a problem.
}
\label{fig:ACpsicontours}
\end{figure}

Finally, the amplitudes of the AC oscillations of $\psi$
are shown in another set of contour plots
in Fig.~\ref{fig:ACpsicontours}.
The observed nodal structures (typical in linear forced systems in 
a bounded domain) show the spatial distribution of 
the modes locked to the driving DC solution.
The magnitude of these AC amplitudes is comparable to  
the DC values in Fig.~\ref{fig:DCpsicontours},
even though the oscillating components of the phases $\phi$ 
are much smaller, of $O(\omega^{-2})$, than their DC values. 
This is consistent with our analysis which assumes
that the mesh current has DC and AC components both of $O(1)$.

\newpage
\section{Parameter region for RS states}
\label{sec:existence}

In this section we determine where in the 
parameter space we expect RS solutions. 
This is a difficult task, partly because the parameter space 
is large. Even after neglecting induced fields 
(i.e.\ $\lambda_\perp=\infty$)
we are left with three parameters: 
$f$, $\Gamma$, and $I_{\mbox{\scriptsize dc}}$.
In addition, there can be multiple attractors coexisting for a given
parameter set. Recall, for example, how
in the previous section Patterns 1, 2 and 4
were obtained using an identical parameter set,
and Pattern 3 also used a similar 
$I_{\mbox{\scriptsize dc}}$ value.
A thorough determination of the parameter regime would then
require a rigorous study of the bifurcations of the branches of all the 
attractors --- an exploration which exceeds the scope of this article,
and is perhaps too detailed to justify the necessary effort.
Here, we take a more heuristic approach, and make several assumptions
to estimate the current interval
$[I^{\mbox{\scriptsize min}}, 
I^{\mbox{\scriptsize max}}]$
in which a given RS state is an attractor,
as a function of $f$ and $\Gamma$.
We base our assumptions on the results of previous sections,
and we demonstrate their validity by additional calculations in the following.

\subsection{Upper current limit}
\label{sec:upperlimit}

We first estimate the upper current $I^{\mbox{\scriptsize max}}$ 
at which a given RS state ceases to be an attractor.
Our {\em first assumption} states that this upper limit
is reached when vortices enter any of the $Q$ regions from the edge.
The entrance of flux might produce further switching of rows 
(resulting in another RS state where the original $Q$ region has been 
subdivided), or a more complicated state where the flux remains static or
moves through the original $Q$ region in a highly nonlinear motion.
In either case, the original RS state is no longer maintained.
As discussed in Sec.~\ref{sec:analysis},
each $Q$ region is decoupled up to the DC leading order
and is equivalent to an isolated superconducting array 
of the same dimensions.
If, as we assume, no vortex has been trapped beforehand in the $Q$ regions,
arrays with more rows depin at smaller values of
$I_{\mbox{\scriptsize dc}}$, as can be shown numerically~\cite{2Dsc}.
Therefore, our {\em second assumption} is that,
as $I_{\mbox{\scriptsize dc}}$ is raised in an RS state,
a vortex first enters the {\it largest} of the remaining $Q$ regions,
causing further break-up of the array.

Thus, once the depinning current values 
for no-vortex superconducting state of any number of rows is known,
these two assumptions enable us to estimate the upper 
$I^{\mbox{\scriptsize max}}$ limit for any given pattern.
For example, Pattern 3 ($S= \{ 4 \}$) has two $Q$ regions 
of the same size (3 rows).
We expect then that this state is not sustainable beyond
the depinning current of a $31 \times 3$ array.
At zero temperature and without disorder,
the likely scenario is that flux enters the center row of each of the 
two regions, 
so that a new RS state, Pattern 2 ($S =\{ 2,4,6 \}$), ensues.
This state has now four $Q$ regions, each consisting of one row.
The upper $I_{\mbox{\scriptsize dc}}$ value for this state should coincide
with the depinning current of the $31 \times 1$ ``ladder'' array.
Beyond this value all rows switch and  Pattern 1 is obtained.
We have indeed observed such a sequence of row-switching events
when we gradually increase $I_{\mbox{\scriptsize dc}}$ from zero,
using a clean initial condition: $\phi=\dot{\phi}=0$ everywhere.
Similarly, the largest $Q$ region in Pattern 4 has 2 rows.
Therefore, $I_{\mbox{\scriptsize dc}}$ should coincide with the 
depinning current of a superconducting no-vortex $31 \times 2$ array.
In Table~\ref{tab:numerics} we summarize the excellent quantitative
agreement between the numerically observed 
$I^{\mbox{\scriptsize max}}$ 
values of several RS patterns, and the depinning currents of 
superconducting arrays with
the same dimensions as their largest $Q$ region.

\begin{table}

\begin{center}

\begin{tabular}{|cc|cc|cc|}  \hline
$S$ &
(Fig.~\protect{\ref{fig:patterns}}) &
$I^{\mbox{\scriptsize min}}$&
$I^{\mbox{\scriptsize max}}$&
$I^{\mbox{\scriptsize dep}}$&
$(N^x \times N^y)$ \\
\hline
\hline

$\{1,\ldots,7\}$ &(\#1)& 
$0.335$ 
&  ---
&  --- & \\

$\{2,4,6\}$& (\#2)& 
$0.335$ & $0.945$ & 
$0.947$ & $(31 \times 1)$\\

$\{4\}$ & (\#3)&
$0.315$ & $0.815$ &  
0.825 & $(31 \times 3)$\\

$\{2,3,4,7\}$ &(\#4)& 
$0.328$ &  $0.912$&
0.912 & $(31 \times 2)$\\

&&&&& \\

$\{1\}$& &
$0.305$ & $0.625$ &
0.622 & $(31 \times 6)$\\

$\{2\}$& & 
$0.305$ & $0.685$ &
0.681 & $(31 \times 5)$\\

$\{3\}$& & 
$0.315$ & $0.776$ &
0.778 & $(31 \times 4)$\\

$\{4\}$ & (\#3)&
0.315 & 0.815 &
0.825 & $(31 \times 3)$ \\
\hline
\end{tabular}

\end{center}

\caption{
Stability intervals 
$[I^{\mbox{\scriptsize min}},
I^{\mbox{\scriptsize max}}]$ (two middle columns) for 
eight RS patterns (two of them identical) in the $31 \times 7$ array
using $f=0.05$ and $\Gamma=0.2$.
The set $S$ denotes the switched row numbers,
and Patterns from Fig.~\protect{\ref{fig:patterns}} are labeled.
The intervals are calculated numerically
by gradually changing $I_{\mbox{\scriptsize dc}}$
and following the corresponding branch of the RS state
until instabilities appear. 
For example, Pattern 2 is found in the interval [0.335, 0.945],
for this set of parameters $(f=0.05,\Gamma=0.2)$.
The upper limit $I^{\mbox{\scriptsize max}}$ can
be predicted accurately by the depinning current 
$I^{\mbox{\scriptsize dep}}$ of
the largest $Q$ region of each pattern 
(with dimensions $N^x \times N^y$, shown in parentheses).
The lower limit $I^{\mbox{\scriptsize min}}$ is harder to estimate,
but the retrapping current $I^{\mbox{\scriptsize ret}}$
of a {\em single} junction serves as a rough estimate:
for $\Gamma =0.2$, the value is $I^{\mbox{\scriptsize ret}}=0.252$,
which is smaller than the observed 
$I^{\mbox{\scriptsize min}}=0.305$--$0.335$.
The first four rows show Patterns 1--4 from 
Figure~\protect{\ref{fig:patterns}}. 
The next four patterns all have a single S row,
but its location is different.
Among these four,
Pattern 3 has the widest stability interval
because its largest Q region ($31 \times 3$) 
has the smallest number of rows.}
\label{tab:numerics}
\end{table}

We have also tested our assumptions with four additional patterns,
all with only one switched row: $S=\{4\}$ (the symmetric Pattern 3), 
$S=\{3\}$, $S=\{2\}$, and $S=\{1\}$ (the most asymmetric pattern).
This illustrates the dependence of 
the upper $I^{\mbox{\scriptsize max}}$  
not on the number of switched rows, as above, 
but on their {\it location}.
For given $f$ and $\Gamma$,
$I^{\mbox{\scriptsize max}}$ 
becomes smaller as the switched row is shifted from 
the middle of the array to the bottom
because the largest $Q$ region increases its size from
$31 \times 3$ to $31 \times 6$.
Excellent agreement is again obtained between
our criterion and the numerical observations~(Table~\ref{tab:numerics}).
\\

We now make the {\em third assumption}
that enables us to estimate 
$I^{\mbox{\scriptsize max}}$ analytically in some cases.
We propose that,  as the drive increases, 
just before the entrance of a vortex into a $Q$ region,
a junction barely holds itself at a critical angle 
\begin{equation}
  \phi^{\mbox{\scriptsize crit}} = \pm \pi/2.
\label{crossingcond}
\end{equation}
When it is forced to turn beyond that value,
depinning takes place,  just as it would  if the junction
were uncoupled. Recall that the
single uncoupled junction under an increasing drive
becomes unstable through a saddle-node bifurcation 
at $I_{\mbox{\scriptsize dc}}=1$,
with $\phi=\pi/2$ as the bifurcation angle.
Although the criterion for global depinning is different in a coupled
array, 
this simple heuristic criterion has been used to predict
the depinning current in ladder arrays
with remarkable accuracy~\cite{ladderdep}.

Take, for instance, an array at zero temperature with small $N^y$ 
in a ground state with no pre-trapped vortices. 
Then, the first junction to cross 
$\phi^{\mbox{\scriptsize crit}} = \pm \pi/2$
is, for $f>0$ and $I_{\mbox{\scriptsize dc}}>0$, 
the vertical junction 
who sits in the center row at the left edge.
Thus, the flux would penetrate the array through that junction and
destroy the RS state. This is readily deduced from  
the circulating current shown in Sec.~\ref{sec:numerics}
which reinforces the drive near the left boundary.
Such a current is due to the presence of the left and right boundaries,
which our large aspect-ratio approximation neglected.
A full analysis of the skin layers would be needed for a general
analytical prediction,
but there are two tractable limiting 
cases of interest.

The first case is a ``small aspect-ratio'' superconducting region,
i.e.\ with many more rows than columns ($N^y \gg N^x$) . 
As discussed in Appendix~\ref{sec:smallaspectratio},
a bulk approximation can then be used,
which approximates accurately the phases near the left and 
right edges --- because, in this case, the skin layers are located
near the top and bottom boundaries.
Should such a region be present in a RS state as a $Q$ region, it 
would be very easily broken even with a small value of  
$I_{\mbox{\scriptsize dc}}$.
In fact, $I^{\mbox{\scriptsize max}}$ can be 
quickly estimated as the drive for which the central 
leftmost vertical junction
crosses the critical angle~(\ref{crossingcond}).
As shown in (\ref{smallarQ}), this junction has indeed the
largest angle.
The depinning value for such a small aspect-ratio region is estimated to be
\begin{equation}
  I^{\mbox{\scriptsize max}}_{\mbox{\scriptsize SAR}} = 
    \frac{1}{2 (N^x+1)}
  \left \{
    1 + \frac{\sin \left[\pi f (2 N^x +1) \right ]}{\sin (\pi f)} 
  \right \} .
\label{smallaruppercurrentlimit}
\end{equation}
From~(\ref{smallaruppercurrentlimit}), the region 
remains stationary when   
$I_{\mbox{\scriptsize dc}} < 
I^{\mbox{\scriptsize max}}_{\mbox{\scriptsize SAR}}$
and $f<1/2N^x$.
If $f>1/2N^x$, a vortex enters the $N^y \gg N^x$ region for any 
$I_{\mbox{\scriptsize dc}} > 0$. 
We have tested these conclusions
numerically with good agreement.
Moreover, note that other physical arguments~\cite{scbooks}
predict that the edge barrier for the penetration of flux 
in this limit would be roughly given by $f_{\rm c} \sim 1/\pi N^x$.
The condition (\ref{smallaruppercurrentlimit}) results from the instability 
of a {\em static} state, and it does not depend on $\Gamma$,
the damping coefficient.
 
\begin{figure}[tbp]
\centerline{\psfig{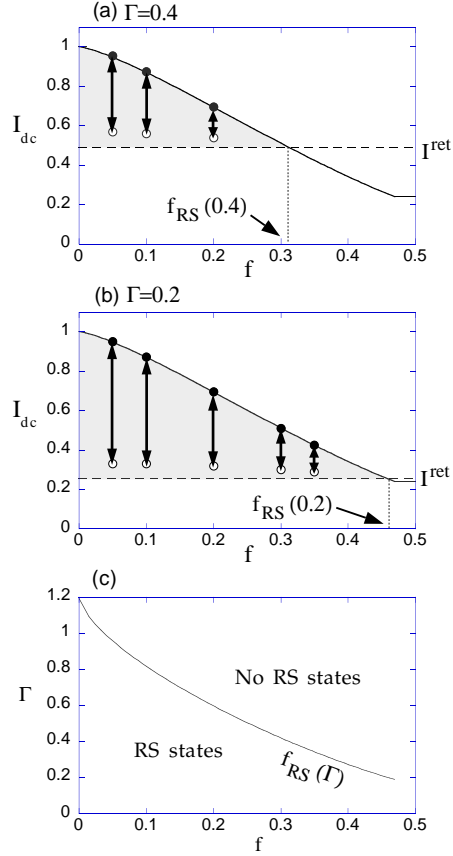}}
\vspace{5mm}
\caption{
(a) Stability region for Pattern 2 with $\Gamma =0.4$.
For $f=0.05, 0.1, 0.2$,
we sweep $I_{\mbox{\scriptsize dc}}$ to determine
numerically the stability interval
$[I^{\mbox{\scriptsize min}},I^{\mbox{\scriptsize max}}]$.
denoted by the vertical arrows with endpoints at
$I^{\mbox{\scriptsize min}}$ ($\circ$) and
$I^{\mbox{\scriptsize max}}$ ($\bullet$).
The solid line is our estimate for 
$I^{\mbox{\scriptsize max}}(f)$,
given by the depinning current 
$I^{\mbox{\scriptsize dep}}$ of a ladder array.
The dashed line is an estimate for
$I^{\mbox{\scriptsize min}}$,
given by the retrapping current $I^{\mbox{ret}}$
of the single junction at $\Gamma=0.4$.
Therefore, the shaded section is the estimated
region of the $I_{\mbox{\scriptsize dc}}$--$f$ plane
where Pattern 2 exists, for $\Gamma =0.4$.
Note that the region does not extend beyond a critical 
$f=f_{\mbox{\scriptsize RS}}(0.4)$.
(b) Same as (a) but for $\Gamma =0.2$.
Although the upper estimate $I^{\mbox{\scriptsize dep}}$ is unchanged,
the lower estimate $I^{\mbox{ret}}$ decreases 
with $\Gamma$. Consequently, Pattern 2 is expected to be observed 
in a larger parameter region for smaller $\Gamma$,
as shown by the five intervals (arrows) obtained numerically.
The region does not extend for $f$ larger than 
$f_{\mbox{\scriptsize RS}}(0.2)$.
(c) Phase diagram for the existence of RS states in the $f$--$\Gamma$ 
parameter plane. 
The curve $f=f_{\mbox{\scriptsize RS}}(\Gamma)$ separates the regions
in which no RS states may or may not appear.
For $\Gamma \geq \Gamma^* \approx 1.2$, 
$I^{\mbox{ret}}=1$, thus, no RS states are expected
for any $f$.
This diagram explains the previous (qualitative) observation that 
RS states occur only when junctions are underdamped and
the applied magnetic field is small.
}
\label{fig:laddercriticalcurve}
\end{figure}

The second case is the ``ladder array'', 
with $N^x$ columns and a single row~\cite{Kimladder}.
Its superconducting states, including states with trapped vortices,
and their bifurcations have been studied comprehensively~\cite{ladderdep}.
One of the results of that work is the curve of the depinning current  
as a function of $f$, shown as a solid line
in Fig.~\ref{fig:laddercriticalcurve}(a) and (b).
This monotonically decreasing curve is again independent of $\Gamma$,
and becomes insensitive to $N^x$, as soon as $N^x$ is greater than about 5.
For $f$ up to about 0.46, 
the depinning is caused by the disappearance of the no-vortex
solution. This part of the curve is well approximated by
the solution of the following implicit nonlinear
equation~\cite{ladderdep}
which comes from imposing (\ref{crossingcond}) to the leftmost junction:
\begin{equation}
\arcsin(1-I^{\mbox{\scriptsize max}}_{\mbox{\scriptsize LAD}}) +
\frac{r-1}{2r} \arccos 
(I^{\mbox{\scriptsize max}}_{\mbox{\scriptsize LAD}})
= \pi f 
\label{ladderdepcond}
\end{equation}
with $r=\alpha + \sqrt{\alpha^2 -1}$
and $\alpha = 1 + \sqrt{1-{
I^{\mbox{\scriptsize max}}_{\mbox{\scriptsize LAD}}}^2}/\cos \pi f$.
After a crossover at $f \sim 0.46$,
the static checkerboard pattern becomes more robust,
and this formula ceases to be valid.
We will not discuss RS states in this high range of $f$.
If our assumptions are correct,
this critical curve should predict the
$I^{\mbox{\scriptsize max}}$ of Pattern 2.
In addition to the single comparison presented in Table~\ref{tab:numerics}
for this pattern, we test it
for $\Gamma=0.2$ and $0.4$ and several values of $f$ 
in Fig.~\ref{fig:laddercriticalcurve}.
As shown there, the numerical $I^{\mbox{\scriptsize max}}$ values 
of Pattern 2 from simulations are predicted very accurately
by the analysis of the ladder depinning point.

Up to now, we have assumed that the magnetic flux penetrates
the Q regions from the left edge of the array.
However, the flux can also enter the array from the top or bottom
boundaries of a Q region in certain situations.
Consider a Q region with a large aspect-ratio and no trapped vortices,
but when the number of rows, say $N^y$, is {\it large}.
In this case, the bulk approximation
obtained in Sec.~\ref{sec:largeaspectratio} can still be used.
From (\ref{largeaspectratiophix})
the maximum angle for the horizontal junctions is
$\phi^x = N^y \pi f$
attained at the top and bottom edges of the region.
It is clear that this value becomes larger than
the critical angle~(\ref{crossingcond}) when $f > 1/2 N^y$.
Thus, for a fixed $N^y$ while $f$ is increased,
the flux would enter the Q region
roughly above that value of the frustration.
The entrance of flux in this manner puts a limit
on the applicability of our analysis.
The assumed {\it no-vortex} Q region is expected to exist only when
the number of rows is smaller than about $1/2f$.
Thus, our analysis does not apply for the initial stages of
the row-switching cascade in large arrays,
when there are still Q regions with many rows.
However, even in such arrays, later steps of the cascade
(when the Q regions have been subdivided)
can be described by assuming no-vortex Q regions.
In addition, our preliminary simulations indicate that
keeping a vortex trapped in a Q region becomes more difficult
both in the presence of $I_{\mbox{\scriptsize dc}}$ 
(which tends to expel the fluxoids from the Q regions), 
and of self-fields (which tend to shield the Q regions from 
the entrance of vortices). 

\subsection{Lower current limit}
\label{sec:lowerlimit}

The parameter regime for the existence of RS states
is also affected by the damping parameter $\Gamma$, as shown in experiments 
and simulations where RS states appear only when the junctions 
are underdamped. However, all the critical currents calculated until 
now are {\it independent} of $\Gamma$. 
We claim that the explanation of the $\Gamma$ dependence of
the RS states requires an estimate of the lower limit 
$I^{\mbox{\scriptsize min}}$.
Unlike the upper limit, in which the superconducting solution
of a $Q$ region ceases to exist,
our numerical observations suggest that 
the lower limit is caused by an instability mechanism
in an $S$ region.
As the bias current is decreased from the values in which
a clean periodic RS state is observed,
S regions start to have trouble in maintaining 
fast whirling oscillations.
Typically, the system begins to show amplitude modulations 
in a slow time-scale, becomes highly nonlinear, 
or gets retrapped altogether.

The variety of possible scenarios makes an accurate estimate much
harder than for $I^{\mbox{\scriptsize max}}$.
In order to make progress,
we have to rely on a rather rough estimate,
based on the dynamics of a single junction.
Recall that the vertical junctions in an $S$ row
are in the resistive (whirling) state. 
For a single underdamped junction, its inertia is enough to maintain a 
whirling solution until very close to the retrapping current
$I^{\mbox{\scriptsize ret}}$,
when it jumps back to the stationary state.
Only near that value does a strong nonlinearity come into play.
Ignoring the inter-junction coupling, we use this current as our 
estimate for the lower limit 
$I^{\mbox{\scriptsize min}}$ of an RS state.
Because of collective effects, the state may not be
immediately retrapped into a stationary state, or, on the contrary, be
retrapped earlier.
However, we expect that, as $I_{\mbox{\scriptsize dc}}$ 
is lowered toward the $I^{\mbox{\scriptsize ret}}$ value,
some nonlinear effects start to become apparent,
so that the simple periodic RS state is altered.

The estimate of $I^{\mbox{\scriptsize ret}}$
is standard~\cite{steve}.
For the underdamped case, 
(i.e., $\Gamma < \Gamma^* \approx 1.2$), 
the retrapping is produced through a homoclinic bifurcation
at $I^{\mbox{\scriptsize ret}}<1$,
and the $I$--$V$ of the single junction is hysteretic.
For all $\Gamma > \Gamma^*$,
$I^{\mbox{\scriptsize ret}}=1$, and there is no hysteresis.
In general, $I^{\mbox{\scriptsize ret}}$ is calculated numerically,
but an asymptotic expression,
$I^{\mbox{\scriptsize ret}} \sim 4 \Gamma/\pi$
can be used as $\Gamma \rightarrow 0$.

From the definition of $I^{\mbox{\scriptsize ret}}$, 
our estimate for $I^{\mbox{\scriptsize min}}$
is thus independent of $f$, $N^x$, $N^y$, 
and the particular RS pattern, but depends on the damping $\Gamma$.
The estimates for $\Gamma=0.2$ and 0.4 are shown
as dashed lines
in Fig.~\ref{fig:laddercriticalcurve}(a) and (b), respectively.
The comparison with the numerical values of
$I^{\mbox{\scriptsize min}}$
(the point when the RS states lose their whirling character)
is not so good, as expected. 
However, our estimate seems to serve as a reasonable first guess.

\subsection{$f$--$\Gamma$ parameter region for RS states}
\label{sec:regionRS}

In the usual experimental setup, the $I$--$V$ characteristic 
of an array is measured by sweeping
the DC current under a constant applied magnetic field
at a fixed temperature 
(which controls the penetration depth $\lambda_\perp$ 
and damping $\Gamma$).
For some combinations of the experimental variables (magnetic field and
temperature) and, thus, of the underlying parameters 
$f$, $\Gamma$, and $\lambda_\perp$,
the $I$--$V$ shows RS steps. 
For others, it does not. 
We will now summarize the preceding sections
and combine their results to estimate 
the ($\Gamma$,$f$) parameter region,
in the limit $\lambda_\perp=\infty$,
in which RS states appear.

First, in Sec.~\ref{sec:upperlimit} 
we showed two limiting cases in which
$I^{\mbox{\scriptsize max}}$ can be obtained analytically, i.e.
when $N^y=1$ (ladder) and when $N^y \gg N^x$.
Numerical simulations~\cite{stroud,ladderdep,2Dsc} show that
$I^{\mbox{\scriptsize max}}$ changes monotonically
between these two limits, as $N^y$ is varied.
(This result is also expected from physical grounds: 
for fixed $N^x$, the magnetic flux penetrates the array more easily 
as $N^y$ is increased.)
An obvious consequence of this is that
the ladder array has the largest parameter domain
for the no-vortex superconducting state.
Therefore, recalling our link between depinning and row-switching, 
RS states whose $Q$ regions are all ladders,
e.g.\ Pattern 2,
are thought to be the most stable in the same sense.
In other words,
when an isolated ladder of length $N^x$ cannot maintain superconductivity, 
the 2D array of size $N^x \times N^y$ cannot show row-switched behavior.
Consequently, the solid curve in Fig.~\ref{fig:laddercriticalcurve}(a)
not only gives the upper limit $I^{\mbox{\scriptsize max}}$ 
for Pattern 2, but also establishes the critical 
$I_{\mbox{\scriptsize dc}}$, for each $f$,
above which no RS states can be observed.

Second, we concluded in Section~\ref{sec:lowerlimit} that 
the $I^{\mbox{\scriptsize min}}$ of {\it all} RS states 
with damping $\Gamma$ can be estimated 
by the retrapping current of a single junction with the same $\Gamma$.
These are the dashed straight lines 
in Fig.~~\ref{fig:laddercriticalcurve}(a),(b).

Hence, the RS states can only exist in the region contained between these 
upper and lower limits, shown as the shaded area in
Fig.~\ref{fig:laddercriticalcurve}(a).
Those limits intersect at a value $f_{\mbox{\scriptsize RS}}(\Gamma)$
beyond which no RS state is possible.
(See Fig.~\ref{fig:laddercriticalcurve}(a)--(b) for the procedure.) 
For junctions of moderate to large damping 
($\Gamma > \Gamma^* \approx 1.2$),
the dashed line is above the curve,
meaning that RS states are impossible for any $f$.
On the other hand, 
for highly underdamped arrays ($\Gamma < 0.2$), 
the line always remains below the curve;
hence, RS states are possible for any $f$
(although the region of $f$ near 1/2 would need more careful 
consideration). Between these two extremes of damping,
the line intersects the curve 
at the critical value $f_{\mbox{\scriptsize RS}}(\Gamma)$, 
which constitutes a phase boundary in the $f$--$\Gamma$ plane.  
In other words, the parameter plane is divided into two regions 
(RS and no-RS) by the curve $f_{\mbox{\scriptsize RS}}(\Gamma)$
in Fig.~\ref{fig:laddercriticalcurve}(c).
This is in qualitative agreement with previous observations,
and awaits more systematic experimental testing.

\newpage
\section{Summary and open problems}
\label{sec:conclusion}

In this article
we have used a weakly-nonlinear perturbative analysis
to study the row-switching phenomenon and to approximate 
the RS solutions.
For the bulk of the array, we have obtained analytical
expressions for the phase and current variables.
In addition, we have estimated the parameter regime
for their existence.
For this, the consideration of the lateral edges has played 
an important role.
The predicted spatial current distributions and the parameter regime
could serve as a guide for
more systematic experimental studies.
In the rest of this section we briefly state open problems
and possible future directions.

The leading-order solutions show good agreement with the numerics,
but leave one phase per row undetermined.
This is $\delta(j)$ in the large aspect-ratio 
approximation~(\ref{DCassumption1}) and such an arbitrary phase 
is still present in the unapproximated
leading order DC equations,
as discussed in Sec.~\ref{sec:analysis}.
However, the full numerics show that there is a slow drift towards a
specific set of $\delta(j)$ for each pattern.
Several authors \cite{landsberg,filawiesen,marino} have studied
this inter-row phase locking in Pattern 1,
but a satisfying answer is yet to be developed.
The zero-field limit ($f=0$) is an exception in that
exact neutral stability and a family of periodic solutions
can be found \cite{wiesen1,petraglia},
implying that there is no inter-row locking.
On the other hand, a slow drift starts to occur
as $f$ is perturbed away from zero~\cite{shidelta}.
We conjecture that the arbitrary phases should be constrained
by a solvability condition in the higher-order expansions of our analysis,
which is automatically satisfied when $f=0$.
Finding that condition, however,
is likely to be an elaborate task.

Our analysis is based on such simplifications as
zero temperature, no disorder, and no self-fields.
Clearly, the effect of relaxing these assumptions should be 
also investigated.
Thermal noise, self-fields  and inhomogeneities  
alter the switching sequence in simulations of the 
row-switching cascade~\cite{stroud,joelrs,ETthesis}. This might explain the
irregularity of the row-switching order observed experimentally
by Tr\'{\i}as~\cite{ETthesis} 
and Lachenmann {\it et al.}~\cite{lachenmann}.
On the other hand, the directed use of disorder (e.g., by removing
some of the edges in the array) might prove a valuable strategy to enhance
the locking property of the arrays~\cite{oppenSIAM}.
Including inductances would also change the current 
distributions~\cite{dominguez,joelstatic,joelrs}.
Previous work~\cite{joelrs,ETthesis}, and our own preliminary calculations 
including self-inductances, show that
RS states persist at least for small inductances.
Our expansion could be extended to include inductances
and then proceed to describe the modified solutions.
However, qualitatively new phenomena can also arise.
For example, it is known~\cite{ETthesis,ETatASC,ETstep}
that, if any inductance is included in the model, 
a coherent state (dynamical checkerboard pattern)
emerges near $f=1/2$ when the RS states cease to exist.

In this article, we have only considered ``clean'' RS states, formed by
whirling and no-vortex superconducting regions.
Thus, we have assumed that the $Q$ rows do not contain any static vortices.
It is generally expected that the depinning of a $Q$ region would
become easier when it contains a pre-trapped vortex.
Therefore, the existence of states with static vortices probably does
{\em not} affect the critical curve 
in Fig.~\ref{fig:laddercriticalcurve}(c).
However, the question of how the depinning of a static 2D array depends on
various parameters ($\Gamma$, $f$, $N^x$ and $N^y$) is not
fully understood, except in the case of the ladder~\cite{ladderdep},
and requires further scrutiny.
Similarly, the $S$ rows in the RS states were assumed to be
in the whirling (normal resistive) state.
Our simulations sometimes show ``generalized'' RS states
which contain one or more rows that are neither switched nor
quiescent, but ``active''.
The states could be born, for instance, when $I_{\mbox{\scriptsize dc}}$
is increased so that vortices start to enter a $Q$ region but not
strongly enough to switch it.
Junctions in the active rows undergo highly nonlinear oscillations,
and propagating vortices are localized.
These states create additional steps in the $I$--$V$ characteristics
between two RS steps, and are detectable.
Thus, they should be considered
for a comprehensive treatment of row-switching.

Apart from investigating the RS states,
we have introduced in this article
a systematic approach to the analysis of the dynamics of 
2D Josephson arrays.
Unlike 1D arrays, which have already led to a great amount of insight
into important phenomena (such as 
soliton propagation and interaction
in the parallel-connected arrays~\cite{shinyalong,soliton},
or synchronization, clustering, and
magnet-like phase transitions
in the series-connected arrays~\cite{1Dseries,WatSwi97}),
2D arrays have been much harder to analyze.
This is partly due to their network equations being more complicated,
and also to their having a wider variety of solutions.
As our weakly-nonlinear analysis shows,
the difficulty regarding the formulation is reduced
by the compact mesh formalism introduced
in the previous numerical 
studies~\cite{sawada,majhofer,dominguez,joelstatic,joelrs,ETthesis}.
We feel that the transparent form of the mesh equations
has the potential to provide analytical information
in the strongly nonlinear regime.

\begin{figure}[tbp]
\centerline{\psfig{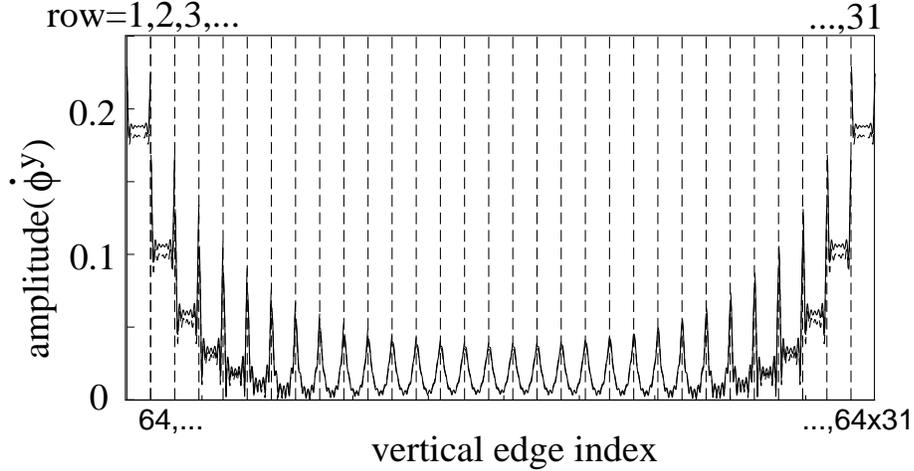}}
\vspace{5mm}
\caption{
Dimensionless AC voltage amplitudes $\dot{\phi}^y$ for Pattern 1
in a large array of size $N^x=63$ and $N^y=31$.
The other parameters are the ones
used in Pattern 1 in Fig.~\protect{\ref{fig:vjcnamp}}.
The amplitudes decay quickly from the boundaries,
and nearly vanish inside the array.
The tendency was already present in Fig.~\protect{\ref{fig:vjcnamp}}
for the $31 \times 7$ array, but it is clearer here.
Consequently, were it to be used as an oscillator, the total AC output
voltage for this pattern would not scale favorably with increasing size.
}
\label{fig:vjcnamp2}
\end{figure}

Of these strongly nonlinear solutions, two are of particular interest.
First, coherent states, such as Pattern 1 in Fig.~\ref{fig:patterns},
might be suitable for oscillator applications,
if $f$ is kept small so that the whole array operates nearly in-phase.
However, for the completely row-switched state to be useful,
overdamped junctions, which rotate less smoothly and, thus, produce
larger AC amplitudes, should be employed~\cite{wiesen1}.
The extension of our analysis for this case 
(concerning only Pattern 1) appears to be straightforward.
However, we can already point out a complication due to the 
spatial distribution of the AC amplitudes.
Recall how the $\dot{\phi^y}$ amplitudes in Pattern 1 
(Fig.~\ref{fig:vjcnamp}) decay from the boundaries
into interior of the array.
This effect is more clearly illustrated in Fig.~\ref{fig:vjcnamp2}
where the AC amplitudes are computed for Pattern 1 
in a larger array ($N^x=63$ and $N^y=31$),
the other parameters being identical.
The amplitudes decay quickly,
and nearly vanish inside the array.
Consequently, the total AC voltage does not increase significantly
even when more junctions are inter-connected.

Finally, flux flow \cite{scbooks,joelrs,ETthesis}
is also a highly nonlinear but disordered regime
in which localized vortices propagate ``diffusively''.
Theoretical studies so far have been based
on phenomenological pictures of vortices and their 
interactions~\cite{vortexparticle}.
A more formal treatment of these solutions 
and a detailed prediction of, for instance, the flux flow resistance
is strongly awaited
both from the theoretical and experimental points of view.

\section*{Acknowledgments}

We thank Steve Strogatz for helpful guidance throughout the course 
of this project. 
Herre van der Zant and Terry Orlando
encouraged our studying the row-switching phenomenon,
and taught us the necessary background on Josephson arrays.
We specially thank Enrique Tr\'{\i}as for sharing with us his
unpublished experimental results and simulations, computer codes, 
and general knowledge of the field.
We also benefited from 
valuable discussions on formulation and computational issues
with Joel Phillips and Jacob White.
Fruitful conversations with Amy Duwel and Mac Beasley 
are also acknowledged.
Although the calculations presented here have been carried out
on standard workstations,
preliminary computations were performed 
at MIT on the Connection Machine 5,
project SCOUT (ARPA contract \#MDA972-92-J-1032).
Research supported in part by NSF grants
DMR-9610042 (through Terry Orlando)
and DMS-9500948 (through Steve Strogatz).
MB gratefully acknowledges support through a Postdoctoral Fellowship
of the Ministerio de Educaci\'on y Cultura of Spain.

\appendix

\newpage
\section{Small aspect-ratio approximation}
\label{sec:smallaspectratio}

Following the large aspect-ratio approximation presented in
Section~\ref{sec:largeaspectratio}, we calculate now a bulk approximation 
to the DC equations~(\ref{DCeqn1})--(\ref{DCeqn4}) 
in a $Q$ region with a small aspect-ratio, i.e. 
when its vertical size $N^y$ is much larger than $N^x$.
Far from the top and bottom boundaries,
the solution is expected to be independent of $j$ 
(assuming there are no trapped vortices).
Then, the DC equations (\ref{DCeqn1}),(\ref{DCeqn2})
simplify to:
\begin{equation}
  \sin \overline{\phi^x_0}(i) = 0 \qquad \mbox{thus} \qquad
  \overline{\phi^x_0}(i) = 0
\label{smallDCsimplified1}
\end{equation}
\begin{equation}
  \overline{\phi^y_0}(i+1) - \overline{\phi^y_0}(i) = -2 \pi f
\label{smallDCsimplified2}
\end{equation}
From (\ref{DCeqn4}) and the boundary conditions~(\ref{DCACBC}) 
we can construct the following telescope sum which must be satisfied
\begin{equation}
  \sum_{i=1}^{N^x+1} \sin \overline{\phi^y_0}(i) = 
  (N^x + 1)\; I_{\mbox{\scriptsize dc}}.
\label{smallDCtelescopesum}
\end{equation}
From these two equations (\ref{smallDCsimplified2}),(\ref{smallDCtelescopesum})
we can then solve for the vertical phases in the bulk of the $Q$ region:
\begin{equation}
  \overline{\phi^y_0}(i) = 2 \pi f \left (\frac{N^x}{2} + 1 -i \right ) 
  + \alpha
\label{smallarQ}
\end{equation}
where 
\[
  \alpha = \arcsin \left\{ 
    \frac{(N^x +1) \sin (\pi f)}{\sin [ (N^x +1) \pi f]}
    I_{\mbox{\scriptsize dc}} 
  \right\} . 
\]
Compare this with the large aspect-ratio case~(\ref{DCsimplified5}) in which
$\overline{\phi^y_0}=\arcsin I_{\mbox{\scriptsize dc}}$ is independent of $i$
in the bulk of a $Q$ region. In contrast, in the present small aspect-ratio
case, the external field $f$ is absorbed now by the {\it vertical} junctions
in order to ensure the flux quantization restriction~(\ref{DCeqn2}).
Note also how, in this case,  consideration of the top and bottom edges,
neglected from the bulk $Q$ region, becomes crucial to introduce
matching across the switched regions or to the array boundaries.
Without the correction from the edges,
phase relations across the $S$ rows are not well defined.
However, the small aspect-ratio approximation is still significant because
it provides a clue to an important question:
what is the lower bound for a $Q$ region to remain unbroken?
Thus, we use this calculation  when we discuss the existence and 
stability of RS patterns in Sec.~\ref{sec:existence}.
In this context, the small aspect-ratio approximation is
the limiting case for which a $Q$ region is most easily broken
by raising either $f$ or $I_{\mbox{\scriptsize dc}}$.



\begin{thebibliography}{99}


\bibitem[*]{SWaddress}
Present address:
Institute of Mathematics \& its Applications (IMA),
University of Minnesota,
Minneapolis, MN 55455; shinya@ima.umn.edu.

\bibitem{herrephasetran} H.S.J.\ van der Zant, L.J.\ Geerligs, and
J.E.\ Mooij, Europhys.\ Lett.\ {\bf 19}, 541 (1992); 
H.S.J.\ van der Zant, F.C.\ Fritschy, W.J.\ Elion,  
L.J.\ Geerligs, and J.E.\ Mooij,
Phys.\ Rev.\ Lett.\ {\bf 69}, 2971 (1992).

\bibitem{geigenlobb}
U.\ Geigenm\"uller, C.J.\ Lobb, and C.B.\ Whan,
  Phys.\ Rev.\ B  {\bf 47}, 348 (1993).

\bibitem{vortexparticle} 
T.P.\ Orlando, J.E.\ Mooij, and H.S.J.\ van der Zant, 
Phys.\ Rev.\ B  {\bf 43}, 10218 (1991);
H.S.J.\ van der Zant, F.C.\ Fritschy, T.P.\ Orlando, and J.E.\ Mooij,
Phys.\ Rev.\ B  {\bf 47}, 295 (1993).

\bibitem{dominguez} 
D.\ Dom\'{\i}nguez and J.V.\ Jos\'e, 
  Phys.\ Rev.\ Lett.  {\bf 69}, 514 (1992); 
D.\ Dom\'{\i}nguez and J.V.\ Jos\'e, 
  Phys.\ Rev.\ B  {\bf 53}, 11692 (1996).

\bibitem{josevortex} 
T.J.\ Hagenaars,  P.H.E.\ Tiesinga, J.E.\ van Himbergen, and J.V.\ Jose, 
Phys.\ Rev.\ B {\bf 50}, 1143 (1994).

\bibitem{wiesen1} 
K.\ Wiesenfeld, S.P.\ Benz, and P.A.A.\ Booi, 
  J.\ Appl.\ Phys.  {\bf 76}, 3835 (1994).

\bibitem{landsberg}
A.S.\ Landsberg, Y.\ Braiman, and K.\ Wiesenfeld, 
  Phys.\ Rev.\ B  {\bf 52}, 15458 (1995).

\bibitem{filawiesen} 
G.\ Filatrella and K.\ Wiesenfeld, 
  J.\ Appl.\ Phys.  {\bf 78}, 1878 (1995). 

\bibitem{oppenlaender} 
J.\ Oppenl\"{a}nder, G.\ Dangelmayr, and W.\ G\"{u}ttinger,
  Phys.\ Rev.\ B  {\bf 54}, 1213 (1996).

\bibitem{marino} 
I.F.\ Marino and T.C.\ Halsey, 
  Phys.\ Rev.\ B  {\bf 50}, 6289 (1994);
I.F.\ Marino, 
  Phys.\ Rev.\ B  {\bf 55}, 551 (1997).

\bibitem{spatiotempchaos}
R.\ Bhagavatula, C.\ Ebner, and C.\ Jayaprakash,
  Phys.\ Rev.\ B  {\bf 45}, 4774 (1992);
  {\em ibid.}, {\bf 50}, 9376 (1994).

\bibitem{trieste95}
  {\it Proc.\ ICTP workshop} (H.A.\ Cerdeira and S.R.\ Shenoy, eds.,
Trieste, Aug.\ 7-11, 1995),
in   Physica  {\bf 222B}, no.4, (1996).

\bibitem{benz}
S.P.\ Benz and C.J.\ Burroughs, 
  Supercond.\ Sci.\ Tech.  {\bf 4}, 561 (1991);
P.A.A.\ Booi and S.P.\ Benz, 
  Appl.\ Phys.\ Lett.  {\bf 64}, 2163 (1994).

\bibitem{shorted2D}
A.\ Petraglia, N.F.\ Pedersen, P.L.\ Christiansen, and A.V.\ Ustinov,
  Phys.\ Rev.\ B  {\bf 55}, 8490 (1997). 

\bibitem{duwel}
A.E.\ Duwel, E.\ Tr\'{\i}as, T.P.\ Orlando, H.S.J.\ van der Zant, 
S.\ Watanabe, and S.H.\ Strogatz,
  J.\ Appl.\ Phys., {\bf 79}, 7864 (1996);
A.E.\ Duwel, S.\ Watanabe, E.\ Tr\'{\i}as, T.P.\ Orlando, 
H.S.J.\ van der Zant, and S.H.\ Strogatz,
  to appear in J.\ Appl.\ Phys.\ (1997).

\bibitem{1Dseries}
P.\ Hadley, M.R.\ Beasley, and K.\ Wiesenfeld,
Phys.\ Rev.\ B, {\bf 38}, 8712 (1988);
S.\ Watanabe and S.H.\ Strogatz,
Physica, {\bf 74D}, 197 (1994);
K.\ Wiesenfeld and J.W.\ Swift,
Phys.\ Rev.\ E, {\bf 51}, 1020 (1995);
K.\ Wiesenfeld, P.\ Colet, and S.H.\ Strogatz,
Phys.\ Rev.\ Lett., {\bf 76}, 404 (1996).

\bibitem{experimrs}
H.S.J.\ van der Zant, C.J.\ M{\"{u}}ller, L.J.\ Geerligs, 
C.J.P.M.\ Harmans, and J.E.\ Mooij,
  Phys.\ Rev.\ B  {\bf 38}, 5154 (1988);
T.S.\ Tighe, A.T.\ Johnson, and M.\ Tinkham,
  Phys.\ Rev.\ B  {\bf 44}, 10286 (1991).

\bibitem{ETthesis}
E.\ Tr\'{\i}as, M.S.\ Thesis (MIT, June 1995);
E.\ Tr\'{\i}as, unpublished.

\bibitem{ETatASC}
E.\ Tr\'{\i}as, M.\ Barahona, T.P.\ Orlando, and H.S.J.\ van der Zant,
IEEE Trans.\ Appl.\ Supercond. {\bf 7}, 3103 (1997).

\bibitem{lachenmann} 
S.G.\ Lachenmann, T.\ Doderer, D.\ Hoffmann, 
R.P.\ H\"ubener, P.A.A.\ Booi, and S.P.\ Benz, 
  Phys.\ Rev.\ B  {\bf 50}, 3158 (1994);
S.G.\ Lachenmann, T.\ Doederer, and R.P.\ H{\"{u}}bener,
  Phys.\ Rev.\ B  {\bf 53}, 14541 (1996).

\bibitem{herreballistic} H.S.J.\ van der Zant, F.C.\ Fritschy, 
T.P.\ Orlando, and J.E.\ Mooij, Europhys.\ Lett.\ {\bf 18}, 343 (1992).

\bibitem{stroud}
W.\ Yu and D.\ Stroud,
  Phys.\ Rev.\ B  {\bf 46}, 14005 (1992);
W.\ Yu, K.H.\ Lee, and D.\ Stroud,
  Phys.\ Rev.\ B  {\bf 47}, 5906 (1993).

\bibitem{channeling} A.\ Brass, H.J.\ Jensen, and A.J.\ Berlinsky,
Phys.\ Rev.\ B {\bf 39}, 102 (1989);
T.\ Matsuda, K.\ Harada, H.\ Kasai, O.\ Kamimura, and A.\ Tonomura, 
Science {\bf 271}, 1393 (1996).
 
\bibitem{highTcSC}
R.\ Kleiner, F.\ Steinmeyer, G.\ Kunkel, and P.\ M\"{u}ller,
  Phys.\ Rev.\ Lett.  {\bf 68}, 2394 (1992).

\bibitem{recentlachenmann}
S.G.\ Lachenmann, T.\ Doderer, R.P.\ Huebener,
T.J.\ Hagenaars, J.E.\ van Himbergen, P.H.E.\ Tiesinga, and J.V.\ Jos\'e
  Phys.\ Rev.\ B {\bf 56}, 5564 (1997).

\bibitem{ETstep}
M.\ Barahona, E.\ Tr\'{\i}as, T.P.\ Orlando, A.E.\ Duwel,
H.S.J.\ van der Zant, S.\ Watanabe, and S.H.\ Strogatz,
  Phys.\ Rev.\ B  {\bf 55}, R11989 (1997).

\bibitem{sawada} 
K.\ Nakajima and Y.\ Sawada, 
  J.\ Appl.\ Phys.  {\bf 52}, 5732 (1981).

\bibitem{chung}
J.S.\ Chung, K.H.\ Lee, and D.\ Stroud, 
  Phys.\ Rev.\ B  {\bf 40}, 6570 (1989).

\bibitem{majhofer} 
A.\ Majhofer, T.\ Wolf, and W.\ Dietrich, 
  Phys.\ Rev.\ B  {\bf 44}, 9634 (1991);
D.\ Reinel, W.\ Dieterich, T.\ Wolf, and A.\ Majhofer,
  Phys.\ Rev.\ B  {\bf 49}, 9118 (1994).

\bibitem{losalamos}
F.\ Falo, A.R.\ Bishop, and P.S.\ Lomdahl,
  Phys.\ Rev.\ B  {\bf 41}, 10983 (1990);
N.\ Gr{\o}nbech-Jensen, A.R.\ Bishop, F.\ Falo, and P.S.\ Lomdahl,
  Phys.\ Rev.\ B  {\bf 45}, 10139 (1992).

\bibitem{joelstatic}  
J.R.\ Phillips, H.S.J.\ van der Zant, J.\ White, and T.P.\ Orlando, 
  Phys.\ Rev.\ B  {\bf 47}, 5219 (1993).

\bibitem{joelrs} 
J.R.\ Phillips, H.S.J.\ van der Zant, and T.P.\ Orlando,
  Phys.\ Rev.\ B  {\bf 50}, 9380 (1994).

\bibitem{petraglia}
A.\ Petraglia, G.\ Filatrella, and G.\ Rotoli,
  Phys.\ Rev.\ B  {\bf 53}, 2732 (1996).

\bibitem{larsen}
B.H.\ Larsen and S.P.\ Benz,
  Appl.\ Phys.\ Lett.  {\bf 66}, 3209 (1995).

\bibitem{likharev} 
K.K.\ Likharev, 
{\it Dynamics of Josephson Junctions and Circuits}  
(Gordon and Breach, New York, 1986).

\bibitem{steve} 
S.H.\ Strogatz,
{\it Nonlinear Dynamics and Chaos: 
with Applications to Physics, Biology, Chemistry, and Engineering}
(Addison-Wesley, Reading, MA, 1994).

\bibitem{scbooks}
T.P.\ Orlando and K.A.\ Delin, 
{\it Foundations of Applied Superconductivity} 
(Addison-Wesley, Reading, MA, 1991);
M.\ Tinkham, 
{\it Introduction to Superconductivity} (2$^{\rm nd}$ ed.) 
(McGraw-Hill, New York, 1996).

\bibitem{choicecaveat}
The choice of the current distribution $I_{\mbox{\scriptsize ext}}$ 
may influence the results when inductance matrices are truncated.
See \protect{\cite{dominguez}}.

\bibitem{strang} 
G.~Strang, 
{\it Introduction to Applied Mathematics}
(Wellesley-Cambridge Press, Wellesley, MA, 1986).

\bibitem{WatSwi97} 
S.\ Watanabe and J.\ Swift, 
J.\ Nonlin.\ Sci.,  to appear (1997).

\bibitem{shinyalong} 
S.\ Watanabe, H.S.J.\ van der Zant, S.H.\ Strogatz, and T.P.\ Orlando, 
Physica {\bf 97D}, 429 (1996).

\bibitem{roots}
The other roots are possible, but this is the root we observe.

\bibitem{BCcaveat}
We note that (\ref{DCswitchedpsi}) ensues 
from the open lateral boundary conditions.
If the right and left boundaries are connected 
to obtain periodic boundary conditions~\cite{stroud,geigenlobb,marino},
the mesh currents $\psi$ do not necessarily vanish 
in the $S$ rows,
and qualitative changes to the results are expected.
We have not studied this case.

\bibitem{shidelta}
S.\ Watanabe, unpublished.

\bibitem{teitel} 
S.\ Teitel and C.\ Jayaprakash, 
Phys.\ Rev.\ B  {\bf 27}, 598 (1983).

\bibitem{2Dsc}
M.\ Barahona, unpublished. See also \protect{\cite{ladderdep}}.

\bibitem{ladderdep} 
M.\ Barahona, S.H.\ Strogatz, and T.P.\ Orlando, 
to appear in  Phys.\ Rev.\ B; 
M.\ Barahona, Ph.D.\ Thesis (MIT, June 1996).

\bibitem{Kimladder}
For ladder arrays with a single {\em column} and $N^y$ rows,
see J.\ Kim, W.G.\ Choe, S.\ Kim, and H.J.\ Lee,
Phys.\ Rev.\ B {\bf 49}, 459 (1994).

\bibitem{oppenSIAM}
J.\ Oppenl\"{a}nder, Poster Presentation,
SIAM Dynamical Systems Conference (Snowbird, 1997).

\bibitem{soliton}
N.F.\ Pedersen and A.V.\ Ustinov,
Supercond.\ Sci.\ \& Tech. {\bf 8}, 389 (1995).

\end{thebibliography}
\end{document}